\documentclass[11pt]{article}

\usepackage{fullpage}
\usepackage{booktabs} %
\usepackage[T1]{fontenc} %
\usepackage{footmisc}

\usepackage[dvipsnames]{xcolor} %
\definecolor{orcidgreen}{HTML}{85A12C} %
\usepackage[hyphens]{url}
\usepackage{hyperref}
\hypersetup{
    colorlinks=true,
    allcolors=Blue,
    linkcolor=Blue,
    citecolor=Blue,
}
\usepackage{microtype} %

\usepackage{amssymb}
\usepackage{amsmath}
\usepackage{amsthm}
\usepackage{orcidlink}
\usepackage{academicons}
\usepackage{xspace}
\usepackage{nicefrac}

\usepackage[nameinlink]{cleveref} %

\newcommand{\calT}{{\mathcal{T}}}

\newcommand{\np}{{{\mathrm{NP}}}}

\usepackage{tikz}
\usetikzlibrary{calc}
\usepackage{subcaption}

\usepackage[square]{natbib}
\usepackage{cleveref}

\theoremstyle{plain}
\newtheorem{theorem}{Theorem}[section]

\newtheorem{proposition}[theorem]{Proposition}

\newtheorem{definition}{Definition}[section]

\newtheorem{remark}{Remark}[section]

\usepackage{tikz}
\usepackage{pgfplots}
\usepgfplotslibrary{groupplots}
\pgfplotsset{compat=1.17,
	legend image code/.code={
		\draw[mark repeat=2,mark phase=2]
		plot coordinates {
			(0cm,0cm)
			(0.15cm,0cm)        %
			(0.3cm,0cm)         %
		};%
}}
\newcommand{\citetapp}[1]{\citet{#1}}

\newcommand{\citeapp}[1]{\cite{#1}}

\newcommand{\posreals}{\mathbb{R}_+}
\newcommand{\katz}{{{\mathrm{katz}}}}
\newcommand{\ged}{{{\mathrm{GED}}}}

\newcommand{\euc}{{{\mathrm{Euc}}}}

\newcommand{\Ord}{{{\mathrm{ord}}}}
\newcommand{\Rps}{{{\mathrm{rps}}}}
\newcommand{\RPS}{{{\mathrm{RPS}}}}
\newcommand{\rt}{{{\mathrm{root}}}}
\newcommand{\linear}{{{\mathrm{lin}}}}

\title{Maps of Tournaments: Distances, Experiments, and Data}
\author{Filip Nikolow\\AGH University, Krak\'ow\\ Poland \and
  Piotr Faliszewski\\AGH University, Krak\'ow\\ Poland  and \and
  Stanis\l{}aw Szufa\\AGH University, Krak\'ow Poland\\
  CNRS, LAMSADE, Universit\'{e} Paris Dauphine -- PSL\\ Paris, France
}

\begin{document}

\maketitle

\begin{abstract}
  We form a ``map of tournaments'' by adapting the map framework from
  the world of elections. By a tournament we mean a complete directed
  graph where the nodes are the players and an edge points from a
  winner of a game to the loser (with no ties allowed).  A map is a
  set of tournaments represented as points on a 2D plane, so that
  their Euclidean distances resemble the distances computed according
  to a given measure. We identify useful distance measures, discuss
  ways of generating random tournaments (and compare them to several
  real-life ones), and show how the maps are helpful in visualizing
  experimental results (also for knockout tournaments).
\end{abstract}

\section{Introduction}

In the map framework, introduced by
\citet{szu-boe-bre-fal-nie-sko-sli-tal:j:map}, we are given a set of
(ordinal) elections together with so-called original distances between
them, and we visualize them as points on a plane so that the distances
between the points resemble the original ones. Such maps are helpful
in presenting experimental results and in understanding relations
between the elections (as well as the random models or the real-life
data sources from which they come). For example, they were
instrumental in better understanding the parameterization of the
classic Mallows
model~\citep{boe-fal-kra:c:mallows-normalization}. Since the inception
of the map idea, it was adapted to
approval elections~\citep{szu-fal-jan-lac-sli-sor-tal:c:map-approval},
stable roommates instances~\citep{boe-hee-szu:j:map-stable-matching},
and voting rules~\citep{fal-lac-sor-szu:c:map-of-rules}. In this paper
we adapt the framework to the case of \emph{tournament graphs}.

Two most prominent types of tournaments studied in AI literature are
the knockout and the round-robin ones. In the former, in each round
the participants are arranged in pairs and play against each other,
with the winners moving up to the next round; in the latter, every
pair of participants plays against each other exactly once.  In
knockout tournaments, the winner is the last player standing, whereas
for round-robin ones there is a whole set of different rules for
victory (e.g., the Copeland rule selects the player(s) who win most
matches).
In either case, for research purposes it is convenient to represent
the results of all possible games---even if not all are actually
played---as a \emph{tournament graph}, i.e., a~complete directed
graph, where for each two players there is an edge pointing from the
winner to the loser (or, from who we expect to win to who we expect to
lose, if perfect information is unavailable;\footnote{It is also
  natural to consider weighted tournament graphs, where the edges are
  labeled with the probability of victory of the more likely winner,
  but we focus on the simpler, deterministic setting.} ties are
forbidden). Such graphs allow us to compute tournament
winners~\citep{fis-bra-har:j:choice-sets,hud:j:tournaments} or
possible
winners~\citep{azi-bri-fis-har-lan-see:j:tournaments-possible-necessary,yan-guo:j:tournaments-possible-winners,bau-hog:c:tournaments-predicting-scheduling},
analyze players' margins of victory
(MoV)~\citep{bri-sch-suk:j:tournaments-mov,dor-pet:c:tournaments-weighted-mov},
check if it is possible to manipulate the
results~\citep{vu-alt-sho:c:tournament-fixing,azi-gas-mac-mat-stu-wal:j:fixing-tournament},
analyze various axiomatic properties of applied solution
concepts~\citep{bra-bri-fis-har:j:tournaments-retentive-sets,bra-bri-har:j:tournaments-extending}
and so on (see the surveys of \citet{bra-bri-har:b:tournaments} and
\citet{suk:c:tournaments} for references going beyond the
just-mentioned examples). Indeed, both knockout and round-robin
tournaments received significant attention within AI and computational
social choice literature, but mostly in terms of theoretical studies,
with experiments gaining traction only recently (in particular, we
mention the works of \citet{sai-suk:c:large-tournament},
\citet{bau-hog:c:tournaments-predicting-scheduling},
\citet{bri-sch-suk:j:tournaments-mov}, and
\citet{fuh-cse-len:c:tournaments-improved-ranking}; works of
\citet{rus:phd:thesis} and
\citet{bra-see:c:tournaments-discriminative} are rare examples of
older experimental studies\footnote{There are experimental works
  beyond AI, e.g., in economics, but we mostly mean those connected to
  algorithmic investigations.}). Our goal is to help researchers
conduct experiments on tournaments in a rigorous, systematic way (we
will often abbreviate \emph{tournament graphs} as \emph{tournaments}).

\paragraph{Components of a Map.}
To form a tournament map, we need three main components: A way to
compute the distances between the tournaments, special tournaments to
mark characteristic spots on the map, and a way to obtain the data.

Regarding the distance measure, an intuitive idea is to use the
\emph{graph edit distance (GED)}: Given two tournaments over the same
number of players, its value is the number of edges that we need to
reverse in one for it to become isomorphic to the
other~\citep{san-sun:j:ged} (the two tournaments may include unrelated
sets of players, such as NBA teams in one and bridge players in the
other, so the distance needs to be invariant to their names; hence the
use of isomorphism). 

As computing GED values is
$\np$-hard~\citep{alo:j:ranking-tournaments}, we also devised a
polynomial-time computable distance, based on the Katz centrality
measure: Given two tournaments, for each of them we compute Katz
centralities of its vertices, put them in a vector (sorted in the
nonincreasing order) and take the $\ell_1$ distance of these
vectors. While this distance and the maps it yields are not as
appealing as the GED ones, they are sufficient
and can be used for larger tournaments.  (Similar type of a distance
was previously used for approval
elections~\citep{szu-fal-jan-lac-sli-sor-tal:c:map-approval}).

With a distance measure in hand, it is also important to identify
characteristic tournaments. Two obvious choices are (a)~the
\emph{ordered} tournament ($T_\Ord$), where the players are ranked
from the strongest to the weakest, and each player defeats all the
weaker ones, and the (b)~\emph{rock-paper-scissors (RPS)} tournament,
where each player defeats half of the other ones (in fact, there are
several nonisomorphic tournaments of this kind; one could also refer
to these tournaments as \emph{regular}).  By mixing them, we obtain an
(approximate) circumference of the tournament space (at least in our
maps).

Finally, to obtain tournaments we either generate synthetic ones
(e.g., using the classic Condorcet noise model or generating
nonisomorphic tournaments), or use real-life data. To this end, we
collected data from NBA seasons and from a number of bridge
tournaments. We find that the NBA tournaments are much closer to
$T_\Ord$ than the bridge ones, but both types of tournaments are quite
specific in that they are far from covering the whole tournament
space. In a number of our experiments, the NBA
tournaments, as well as those nearby, gave the most varied results,
which suggests that they are in a particularly diverse part of the
tournament space.

We have also analyzed a dataset analogous to that of
\citet{bri-sch-suk:j:tournaments-mov} and a dataset derived from the
elections of \citet{fal-kac-sor-szu-was:c:microscope}. The former
contains realistic data, whereas the latter one is quite specific and
sometimes shows unexpected behavior (for example, elections generated
uniformly at random, using the impartial culture model, tend to
generate tournaments with fairly strong structure). Yet, many
tournaments derived from elections land in a similar area of our maps
as the NBA ones, so they can be useful, but should be
analyzed with care.

\paragraph{Evaluation.}
We evaluate our maps in several ways. First, we compute the
correlation between the GED and Katz distances. We find that, overall,
the Katz distance is acceptable (if not ideal). Next, we show where
tournaments generated according to a number of statistical cultures
land on our maps and how it compares to real-life data. Finally, we
show how our maps help in visualizing a number of experiments (in
particular, regarding the number of winners that various solutions
concepts for round-robin tournaments generate, and regarding possible
manipulations of knockout tournaments). Based on such visualizations
one can, for example, identify areas of the tournament space where
various phenomena happen (as we do by pointing to the vicinity of
the NBA tournaments) or predict how structure of the tournaments
affects a particular quantity (e.g., in our experiments the distance
from $T_\Ord$ is a good estimator of the time needed to compute the
tournament's Slater winners using an ILP solver).

\paragraph{Supplementary Material.} For suplementary materials, visit:
\url{https://github.com/Project-PRAGMA/Map-of-Tournaments-ECAI-2025}

\section{Preliminaries}

We use $[n]$ to denote the set $\{1,2,...,n\}$.
By $\posreals$ we mean the set of
nonnegative real numbers.

\paragraph{Graphs and Tournaments.}
A \emph{directed graph} is a pair $G = (V,E)$, where $V$ is the set of
vertices and $E$ is the set of edges (sometimes called arcs in the
literature). Each edge is a pair of vertices, where we interpret
$(v,u)$ as pointing from $v$ to $u$.  By $\delta(v)$ we mean the
outdegree of vertex $v$, i.e., the number of edges pointing out from
it. All our graphs are directed.

A \emph{tournament graph} (or a \emph{tournament}, for short) is a
graph where for each two vertices $u$ and $v$ there is exactly one
edge between them (either from $u$ to $v$ or the other way round).  We
often refer to the vertices in a tournament as the \emph{players} or
\emph{participants} and we interpret an edge $(v,u)$ as meaning that
player $v$ wins against player $u$.  We write %
$\calT_n$ to denote the set of all tournaments over $n$ players.

We say that graphs $G_1 = (V_1,E_1)$ and $G_2 = (V_2,E_2)$ are
isomorphic if there is a bijection $\pi \colon V_1 \rightarrow V_2$
such that for each two vertices $v,u \in V_1$, $G_1$ contains edge
$(v,u)$ if and only if $G_2$ contains edge $(\pi(v),\pi(u))$.

\paragraph{Distances.}
Given a set $X$, a function $d \colon X\times X \rightarrow \posreals$
is a \emph{distance} if for all $x,y,z \in X$ it holds that: (a)~$d(x,y) = 0$
if and only if $x = y$, (b)~$d(x,y) = d(y,x)$, and
(c)~$d(x,y) \leq d(x,z) + d(z,y)$.
If we relax the first condition to require that $d(x,x) = 0$ for all $x \in X$,
then we get a pseudodistance.
Abusing the terminology, we often refer to pseudodistances as distances.

\paragraph{Distances Between Vectors.}
Given two real-valued vectors, $x = (x_1, \ldots, x_n)$ and
$y = (y_1, \ldots, y_n)$, their $\ell_1$ distance is
$\ell_1(x,y) = |x_1-y_1| + \cdots + |x_n-y_n|$.

\paragraph{Distances Between Tournaments.}
Let $d$ be a distance between tournaments (over the same numbers of
players). We say that $d$ is \emph{isomorphism respecting}
(\emph{isomorphic}, for short) if it is invariant to reordering the
players.  So, if $T_1$ and $T_2$ are isomorphic and $d$ is such a
distance, then $d(T_1,T_2) = 0$.

\paragraph{Centrality Measures.}

A centrality measure is a function that given a vertex, associates it
with its ``importance'' (vertex outdegree is among the simplest such
measures).  Let us fix numbers $\alpha \in [0,1)$ and $\beta \geq 0$,
and a graph $G = (V,E)$, with $V = \{v_1,\ldots,v_n\}$.  Katz
centralities of $v_1, \ldots, v_n$, denoted
$\katz(v_1), \ldots, \katz(v_n)$, are numbers such
that~\citep{kat:j:katz}:
\begin{align*}
  \textstyle
  \katz(v_i) = \sum_{(v_i,v_j) \in E} \alpha \cdot \katz(v_j) + \beta. 
\end{align*}
We use $\alpha = 0.1$ and $\beta = 1$.\footnote{These are standard
  settings from the NetworkX Python library.}  Katz centrality is
related to the eigenvector~\citep{bon:j:centralities} and PageRank
centralities~\citep{pag-bri-mot-win} as it implements the same general
intuition: A vertex in a tournament is ``important'' (or, ``strong''
in the language of tournaments) if it connects to other ``important''
vertices (if it beats other ``strong'' players).  For further
discussions of eigenvector and PageRank centralities, see, e.g., their
axiomatic characterizations due to Wąs and
Skibski~\citep{was-ski:c:centrality-katz-eigenvector,was-ski:j:pagerank}.
Katz also has some connection to the outdegree centrality, as its
value most strongly depends on the values of its neighbors.

\section{Generating and Collecting Tournaments}

Next, we describe the tournaments that we present in our maps.
In most of our experiments, we either focus on tournaments with $7$ or
$12$ players, but we also consider other sizes.

\subsection{Characteristic Tournaments}
\label{sec:characteristic-tournaments}
Let $n$ be the number of players. In the ordered tournament, $T_\Ord$,
there is a strict ordering of the players from the strongest to the
weakest and each player wins with all the weaker ones (i.e.,
$T_\Ord$ is transitive; the number of players will always be clear
from the context so we omit it in notation).  Intuitively, $T_\Ord$ is
the ``simplest'' tournament, with a clear winner.

On the other end of the spectrum we have the $\RPS$ family of the
``rock-paper-scissors'' tournaments, where each player wins against
half of the other ones (modulo parity of the number of players) and we
typically expect a multiway tie. The name stems from a popular game,
represented as a three-player tournament, where each player wins
against one opponent and loses to the other one. The $\RPS$ family
includes many nonisomorphic tournaments, of which we are particularly
interested in $T_\Rps$, formed as follows: We let the players set be
$V = \{v_0, \ldots, v_{n-1}\}$ and as~$i$ goes from $0$ to $n-1$, we
add an edge from $v_i$ to each of
$v_{(i+1) \mod n}, \ldots, v_{(i+\lfloor \frac{n}{2} \rfloor) \mod n}$
(unless the result of this game was settled already, which happens for
even numbers of players, once for each of the players
$v_{\frac{n}{2}}, \ldots, v_{n-1}$).

We also consider two families of tournaments that fall between the
above two extremes. In the $\Ord/\RPS$ one, for each $k \in [n-3]$
there is a tournament where $k$ players form the $T_\Ord$
subtournament and $n-k$ form the $T_\Rps$ one, with all players from
$T_\Ord$ defeating all those from $T_\Rps$ (note that we consider $k$
only up to $n-3$, so the $T_\Rps$ tournament has at least $3$
players and, hence, is nondegenerate).  Family $\RPS/\Ord$ is defined analogously, but with the
edges between the $T_\Ord$ and $T_\Rps$ subtournaments reversed.

\subsection{Statistical Models of Tournaments}
\label{sec:statistical-models-of-tournaments}
There are two main ways of generating random tournaments.  We can
either use ``native'' models that generate tournaments directly, or we
can first generate ordinal elections and then derive their majority
relations (the latter approach is taken, among other options, e.g., by
\citet{bri-sch-suk:j:tournaments-mov} and
\citet{bra-see:c:tournaments-discriminative}). We focus on the
following direct models, but we will mention election-related ones
later (let $V = \{v_1, \ldots, v_n\}$ be the set of players).

\begin{description}
\item[Nonisomorphic Sampler.] We use the package \texttt{nauty} of
  \citet{mck-pip:j:nauty} to generate nonisomorphic tournaments. For
  the case of $7$ players, we generate all such tournaments and for
  other cases only a subset of them; \texttt{nauty} does not guarantee
  uniform distribution in any clear sense,
  so we generate a large
  number of such tournaments and sample from them (for improved diversity).

\item[Uniform Model.] In the uniform model, for each pair of players
  we choose the result of their game (i.e., the orientation of the
  edge between them) independently, uniformly at random.

\item[Condorcet Noise Model.] We are given probability $p$ as a
  parameter. We start with the ordered tournament and for each pair of
  players we reverse the result of their game with probability $1-p$.

\item[Strength Models.] Each player $v_i$ has some strength
  $w(i) \in \posreals$. To generate a tournament, for each two players
  $v_i$ and $v_j$, the former wins with probability
  $\frac{w(i)}{w(i)+w(j)}$. E.g., we consider exponential
  strength functions, $w^\alpha_{\exp}(i) = \alpha^i$, where
  $\alpha \geq 1$ is a
  parameter. %

\end{description}

\begin{remark}\label{rem:ic}
  Models similar to the strength one appear in the literature in
  various forms and shapes. In particular,
  \citet{ryv-ort:j:elo-tournament} suggest how players' abilities
  should translate to victory probabilities (see also the work of
  \citet{ald:j:elo-tournament}). \citet{sai-suk:c:large-tournament}
  consider a ``gap model'' based on a similar premise.
\end{remark}

\subsection{Collected Real-Life Data}
\label{sec:collected-real-life-data}

We have collected data from the NBA League (top US basketball league)
and from Polish Sports Bridge Association (which manages the bridge
card game leagues in Poland).

\paragraph{NBA.}
We have formed a tournament for each of the 20 regular NBA seasons
(i.e., not counting playoffs) between the 2000/2001 and 2019/2020
ones.  The first four of these seasons include 29 teams and the later
ones include 30 teams. During each season, each NBA team plays every
other one between two and four times (depending on the conference and
the division they are in).  Given two teams, we convert the results of
their games to an edge in the tournament as follows:
\begin{enumerate}
\item The team that, altogether, scored more points against the other
  one wins.
\item If this leads to a tie, the team with greater point advantage in
  its best game against the other one wins.
\item If this is not decisive, the team with fewer personal fouls in
  all their games against the other one wins.
\item If we still have a tie (which only happened a few times
  throughout all seasons), we choose the result at random.
\end{enumerate}

\paragraph{Bridge.}
Polish Sports Bridge Association stores an archive of the results from
the Polish leagues (\url{https://www.pzbs.pl/liga-archiwum}); we
collected the results of the round-robin parts of the top, north, and
south leagues for seasons from 2017/2018 to 2022/2023. Results of the
matches were directly converted to tournament graphs using the Victory
Points scored in head-to-head contest between each two teams (the team
with higher score wins).  Altogether, this dataset consists of 17
tournaments and each tournament has 16 players.

\begin{figure}[t]
  \centering
  \begin{subfigure}{0.3\textwidth}
    \centering
    \includegraphics[width=0.8\columnwidth]{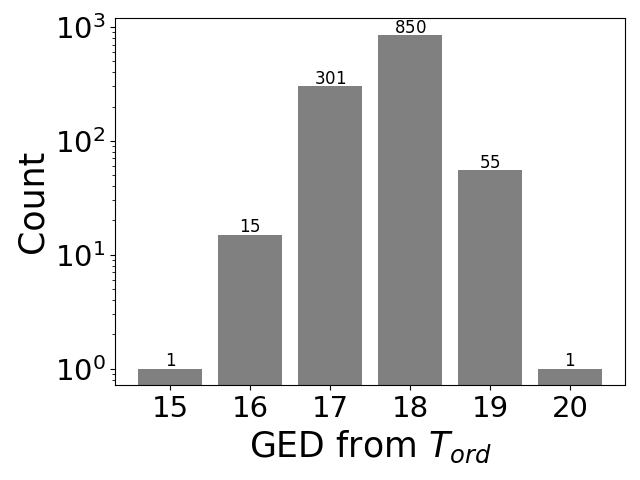}%
    \caption{Histogram of GED distances between $T_\Ord$ and RPS
      ones. \phantom{dwa duze koty miaucza} \phantom{i}}
    \label{fig:rps:histogram}
  \end{subfigure}\quad
  \begin{subfigure}{0.3\textwidth}
    \centering
    \includegraphics[width=0.8\linewidth]{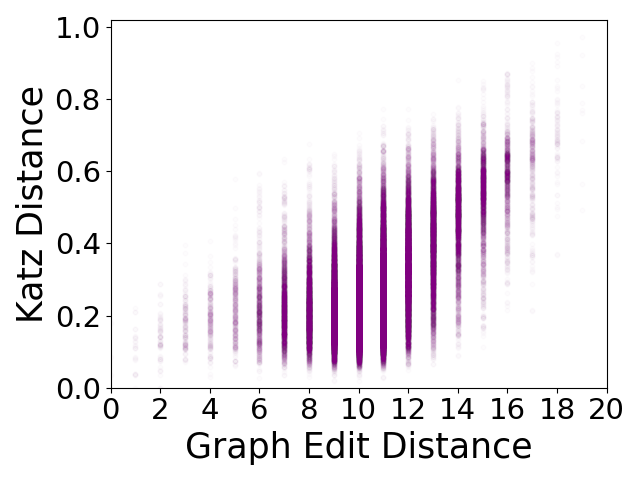}%
    \caption{Comparison of the GED and Katz distances on the Size-12
      dataset. \phantom{trzy niebieskie} \phantom{tygrysy}}
    \label{fig:size12:ged-katz}
  \end{subfigure}\quad
  \begin{subfigure}{0.3\textwidth}
    \centering
    \includegraphics[width=0.8\columnwidth]{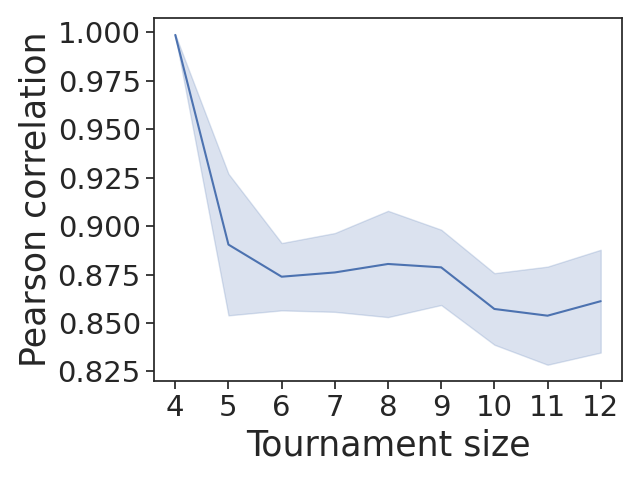}%
    \caption{PCC between  GED and Katz distances, depending
      on the number of players.}
  \label{fig:uniform:ged-katz}
\end{subfigure}
\bigskip

\caption{Basic insights regarding
  tournaments. Plot (a) shows a histogram of GED distances between
  $T_\Ord$ and all RPS tournaments with 11 players, plot (b) compares
  the GED and Katz distances in the Size-12 dataset (each point is a
  pair of tournaments, its $x$ coordinate is their GED distance and
  its $y$ coordinate is the Katz distance), and plot (c) shows PCC of
  the GED and Katz distances between $T_\Ord$ and $100$ tournaments
  generated using the uniform model, depending on the number of
  players (for each number of players we repeated the experiment $10$
  times; the shaded area shows standard deviation).}
  \label{fig:rps-ged-katz}
\bigskip

\end{figure}

\subsection{Dataset Composition}
Finally, we describe four datasets that we consider. Each of them
includes the $T_\Ord$ and $T_\Rps$ tournaments, as well as $n-3$
tournaments from the $\Ord/\RPS$ family and $n-3$ tournaments from the
$\RPS/\Ord$ one (where $n$ is the number of players in the tournaments
from a given dataset) in addition to the tournaments described below.

\paragraph{Size-7 Dataset.}
This dataset consists of all 456 nonisomorphic tournaments over $7$
players (this was the largest number of players for which it was
reasonable to generate all the tournaments; for slightly more players
the number of such tournaments is too large to conveniently present
them on a map, and for significantly more players even computing all
nonisomorphic graphs is challenging).

\paragraph{Size-12 Dataset.}
This dataset regards 12 players and is our main object of study.  We
formed this dataset as follows:
\begin{enumerate}
\item We generated 100 nonisomorphic tournaments.\footnote{We used \texttt{nauty}
    with the res/mod option set to $0/200000$, generated 800 000 nonisomorphic graphs,
    and sampled 100.}
\item We generated 20 tournaments using the uniform model.
\item For each $p \in \{0.1,0.2,0.3,0.4\}$, we generated 5 tournaments
  using the Condorcet noise model.
\item For each function among $w^e_{\exp}(i)=e^i$, $w^2_{\exp}(i) = 2^i$,
  $w_\linear(i) = i$, $w_{\log}(i) = \log(i)$, $w_\rt(i) = \sqrt{i}$, we
  generated $10$ tournaments using the strength model with this
  strength function.  We also generated $10$ tournaments by choosing
  $\alpha \in [1,2]$ uniformly at random (separately for each
  tournament) and using the $w^\alpha_{\exp}(i) = \alpha^i$.
\item For each of the NBA and bridge tournaments we, formed four copies of this
  tournament and restricted each of them to randomly (and independently) selected 12
  players.  Hence, altogether, we have 80 NBA tournaments and 68
  bridge tournaments.

\end{enumerate}

\paragraph{MoV Dataset.}
This dataset consists of tournaments generated in the same way as in
the work of \citet{bri-sch-suk:j:tournaments-mov}, except that we
considered $12$ players (to maintain similarity to the above dataset)
and generated $10$ tournaments for each statistical model instead of
$100$. We also included 100 nonisomorphic tournaments from
\texttt{nauty}, as in the Size-12 dataset.

\paragraph{Election Dataset.}
We generated tournaments based on the election data of
\citet{fal-kac-sor-szu-was:c:microscope}. In their dataset, each
election has $8$ candidates and $96$ voters, where each voter ranks
the candidates from the most to the least desirable one. Given such an
election, we form a tournament where candidates are players and a
candidate $a$ wins against candidate~$b$ if he or she is preferred by a
majority of the voters (i.e., we use their majority relations; we
resolve ties uniformly at random).

\begin{remark}\label{rem:ic}
  In particular, the election dataset includes elections generated using the
  impartial culture model, where each vote (i.e., each ranking of the
  candidates) is sampled uniformly at random. Consequently, given a
  pair of candidates $a$ and $b$, the probability that $a$ would
  defeat $b$ if we first generated an election using the impartial
  culture model and then generated a tournament from this election is
  $0.5$. So, thus generated tournaments seem to be analogous to those
  from the uniform model, but the results of the games are not 
  independent (for example, if we consider three candidates, $a$, $b$, and $c$,
  and three voters, then the probability that in an election drawn using the impartial culture model a majority of the voters ranks $a$ over $b$---under the condition that a majority of the voters also 
  ranks $a$ over $c$---is $\nicefrac{68}{108} \approx 0.629$ and not $0.5$).
\end{remark}
We omit from the election dataset tournaments that are equal to
$T_\Ord$ (for readers familiar with elections literature, this
includes, e.g., tournaments generated from single-peaked elections).
For better comparison with the other datasets, we included $500$
nonisomorphic tournaments from \texttt{nauty}.

\section{Distances Between Tournaments}

Below we describe two distances that we use to generate our maps.  The
first one is precise, but hard to compute for larger graphs, whereas
the latter is fast to compute, but less precise.

\subsection{Graph Edit Distance}

Given tournaments $T_1, T_2 \in \calT_n$, their \emph{graph edit
  distance (GED)}, denoted $d_\ged(T_1,T_2)$, is the number of edges
that we need to reverse in $T_1$ to obtain a tournament isomorphic to
$T_2$~\citep{san-sun:j:ged}.  It is well-known that $d_\ged$ is a
pseudodistance and, by definition, it is isomorphic. On the negative
side, deciding if the GED distance between two tournaments is at most
a given value is $\np$-hard~\cite{alo:j:ranking-tournaments}, but it
can be computed using an ILP formulation of \citet{ler-abu-rav-her-ada:b:ged-blp}. GED is
natural, intuitive, and appealing. Below we provide some of its
features relevant to our analysis.

First, we observe that GED is closely related to the \emph{pairwise}
distance of \citet{szu-fal-sko-sli-tal:c:map}. For our discussion, the
only difference is that the latter is defined on weighted tournaments,
where each edge has a weight between $0$ and $1$ (in the language of
\citet{szu-fal-sko-sli-tal:c:map}, pairwise is defined on
\emph{weighted majority relations} of ordinal elections); GED is equal
to pairwise if we assume all edges to have weight $1$. Consequently,
the largest GED distance between two tournaments is bounded from above
by the largest pairwise distance between two weighted tournaments,
which for tournaments with $n$ players is
$\frac{1}{4}(n^2-n)$~\citep{boe-fal-nie-szu-was:c:metrics}. By the
next proposition, we see that the largest GED distance is comparable
to this value.

\begin{proposition}\label{pro:ord-rps}
  If $n$ is an odd integer representing the number of players, then
  $d_\ged(T_\Ord,T_\Rps) = \nicefrac{1}{8}(n^2-1)$.
\end{proposition}
However, the above is certainly not the largest possible GED value as
even in the $\RPS$ family there are graphs that are further away from
$T_\Ord$ than $T_\Rps$. Indeed, in \Cref{fig:rps:histogram} we see a
histogram of distances between the $\RPS$ tournaments and $T_\Ord$,
for $11$ players (this was the largest size for which we were able to
generate all $\RPS$ tournaments). In this case,
$d_\ged(T_\Ord,T_\Rps) = 15$ and all the other $\RPS$ tournaments are
further away from $T_\Ord$.

\begin{figure}[t]
  \centering
  \begin{subfigure}{0.28\textwidth}
    \includegraphics[width=\textwidth]{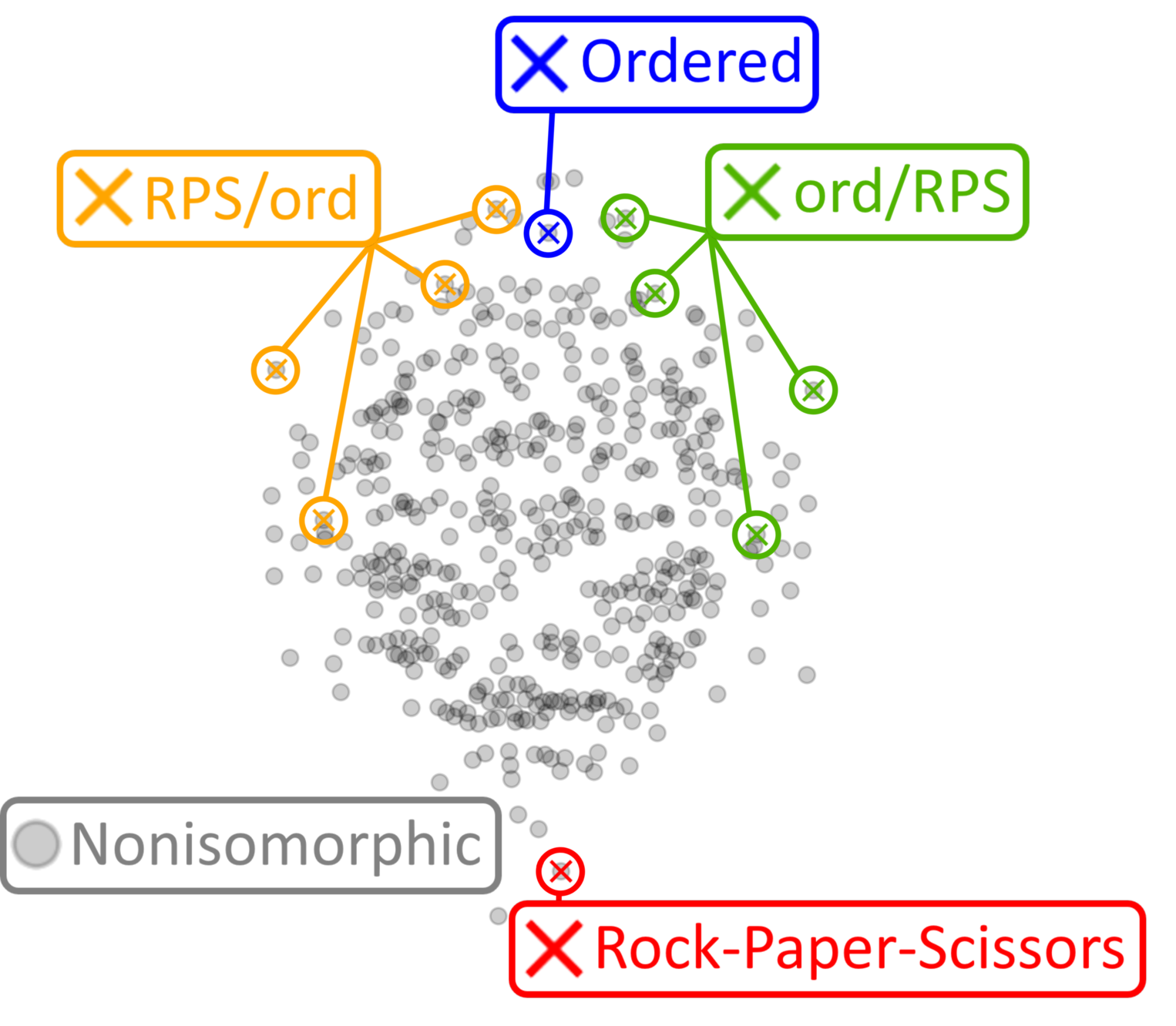}
    \caption{Size-7 Dataset.}
    \label{fig:dataset-7-ged}
\end{subfigure}~
  \begin{subfigure}{0.33\textwidth}
    \includegraphics[width=\textwidth, trim={0cm 0cm 0cm 0cm}, clip]{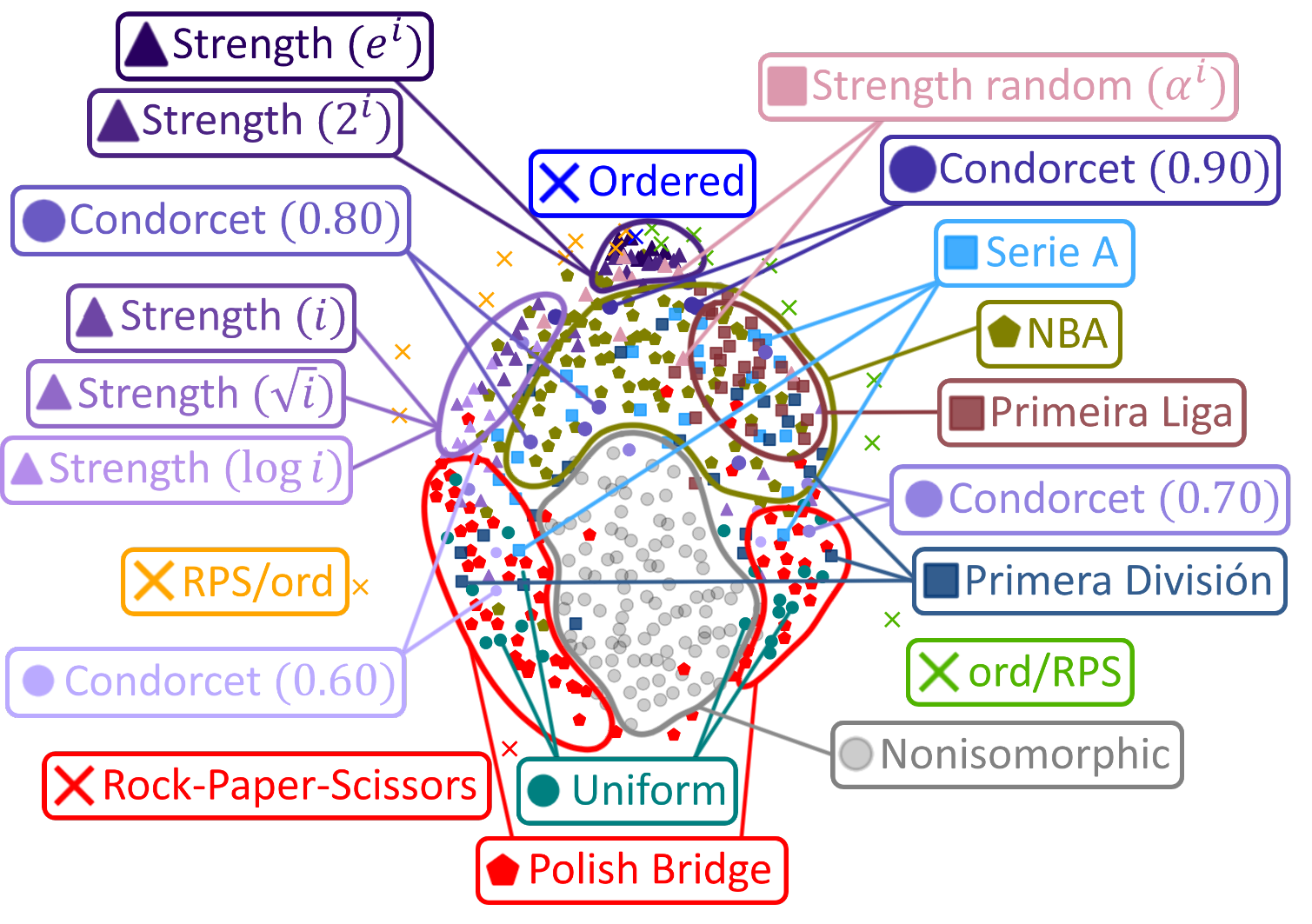}
    \caption{Size-12 Dataset.}
    \label{fig:dataset-12-ged}
  \end{subfigure}~
  \begin{subfigure}{0.33\textwidth}
    \includegraphics[width=\textwidth]{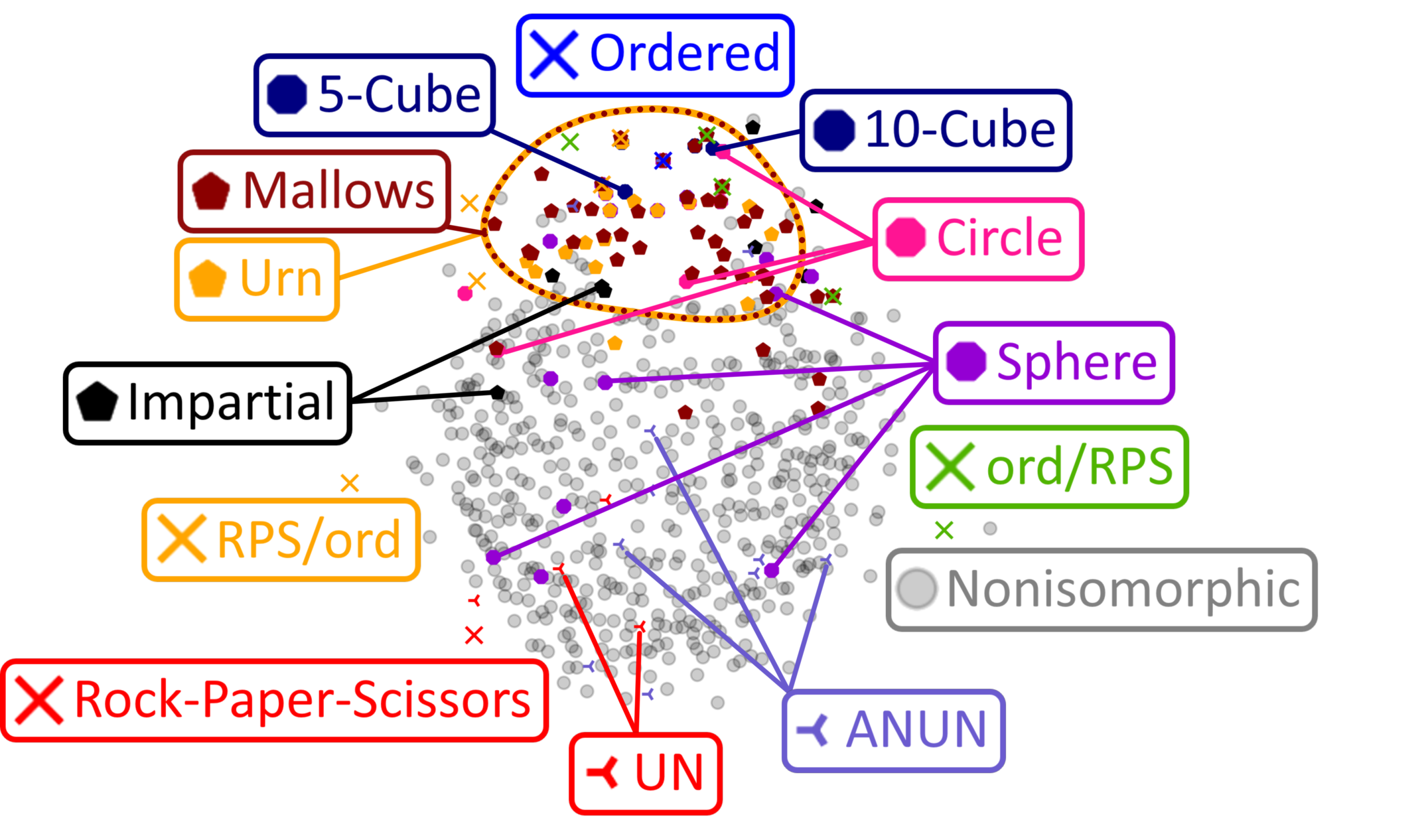}
    \caption{Election Dataset ($n=8$).}
    \label{fig:dataset-election-ged}
  \end{subfigure}~
  \bigskip
  
 \begin{subfigure}{0.28\textwidth}
    \includegraphics[width=\textwidth]{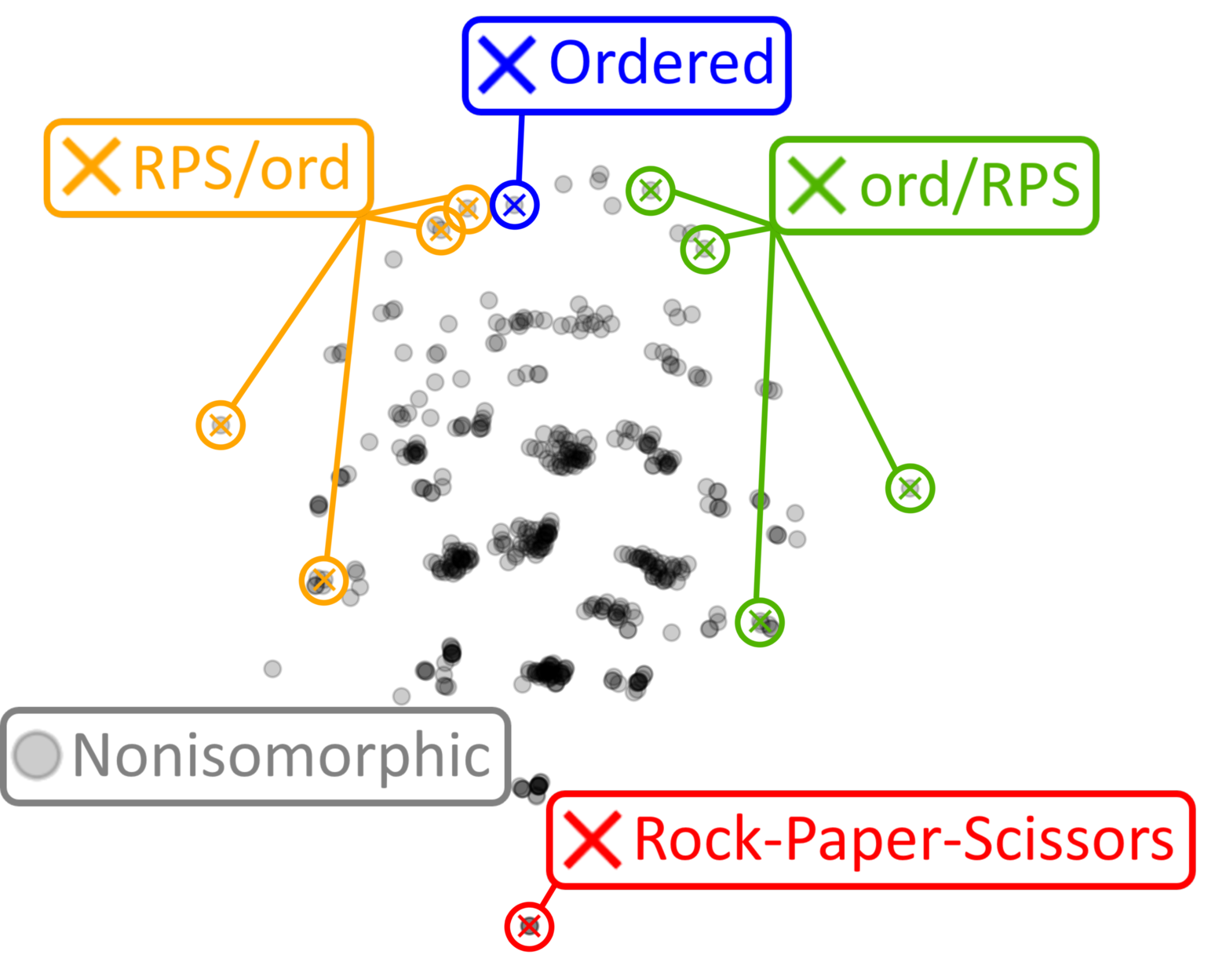}
    \caption{Size-7 Dataset.}
    \label{fig:dataset-7-katz}
  \end{subfigure}
    \begin{subfigure}{0.33\textwidth}
    \includegraphics[width=\textwidth]{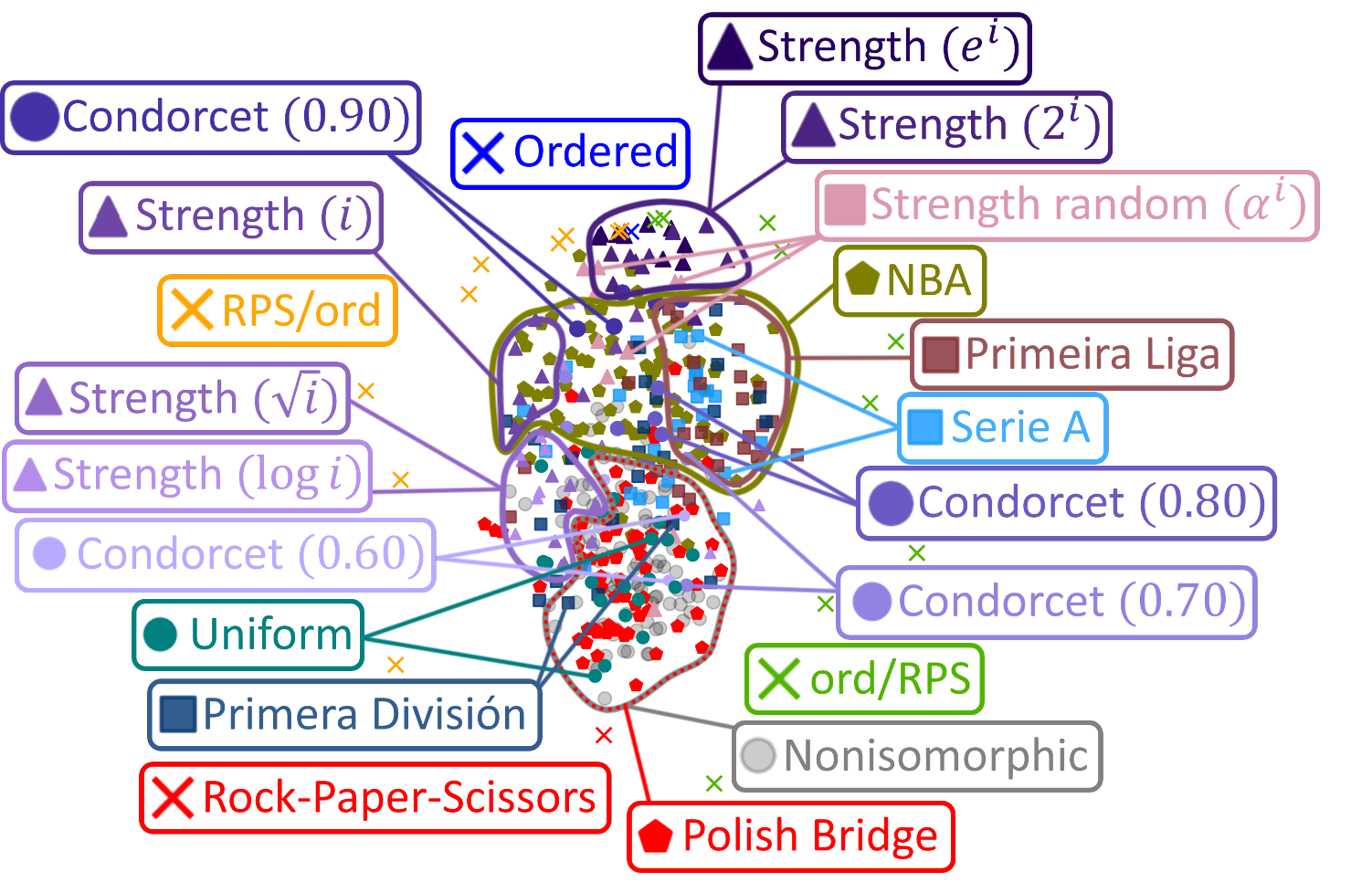}
    \caption{Size-12 Dataset.}
    \label{fig:dataset-12-katz}
  \end{subfigure}
  \begin{subfigure}{0.33\textwidth}
    \includegraphics[width=\textwidth]{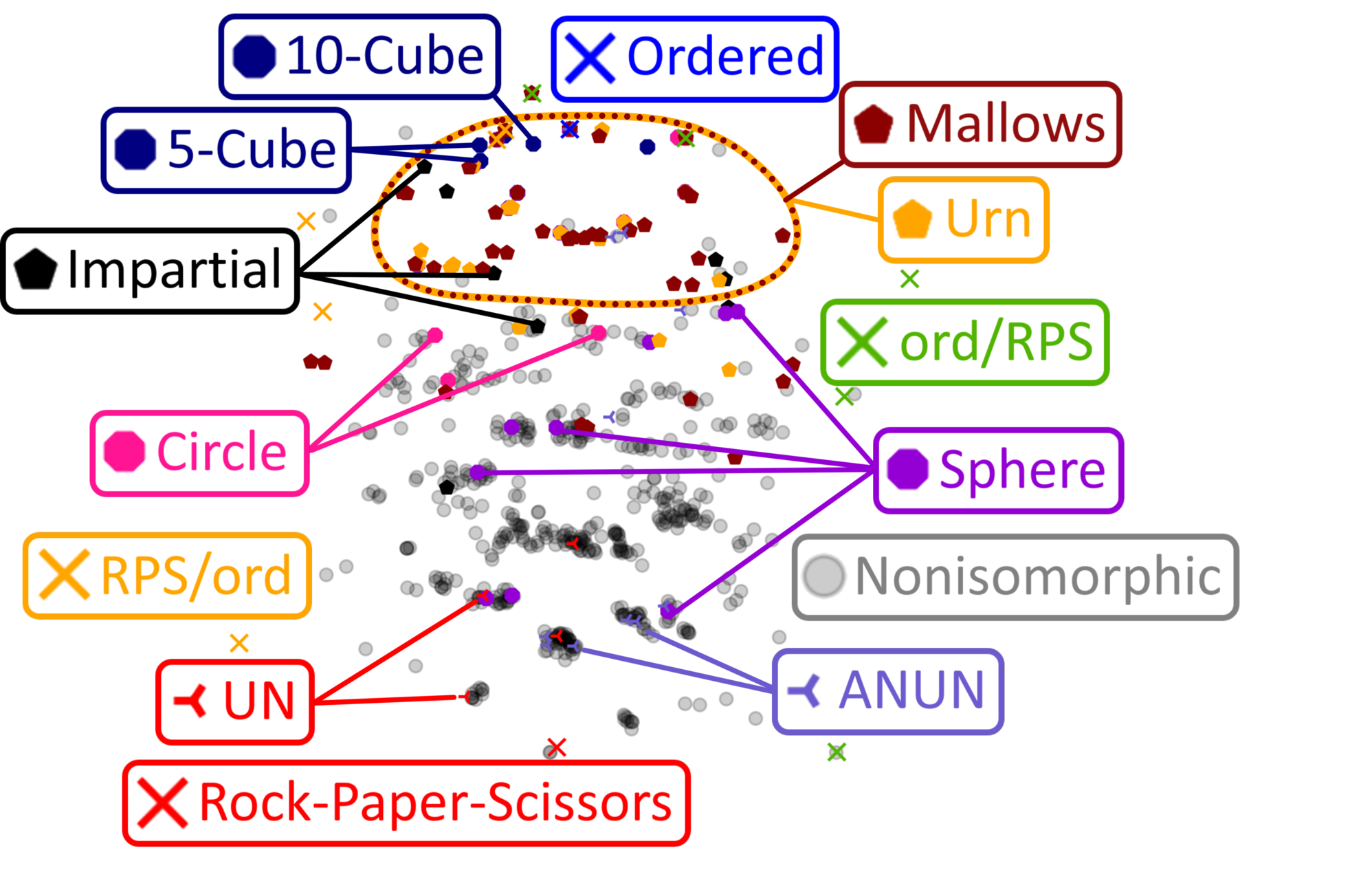}
    \caption{Election Dataset ($n=8$).}
    \label{fig:dataset-election-katz}
  \end{subfigure}
  \bigskip

  \caption{Maps of different datasets computed using GED (upper row) and Katz distance (lower row).}
  \label{fig:map:all_datasets} 
  \bigskip
\end{figure}

It turns out that for sufficiently large $n$, the farthest distance of
a tournament from $T_\Ord$ is of the form
$\frac{1}{4}(n^2+n) \pm O(n^{\frac{3}{2}})$; see sequence A003141 in
the encyclopedia of integer sequences~\citep{oeis}, as well as the
book of \citet[p.~42]{spe-erd:b:probabilistic-method}.  Thus, for large
enough $n$, the largest pairwise distance is not too far off from the
largest GED one. Nonetheless, if needed, we prefer to normalize GED
distances by the distance between $T_\Ord$ and $T_\Rps$ because the
largest possible distance is at most twice as large and value
$d_\ged(T_\Ord,T_\Rps)$ has clear interpretation. In our data
we did not observe too many tournaments that would be much further
away.

\begin{figure}[t]
  \centering
  \begin{subfigure}{0.45\textwidth}
    \includegraphics[width=\textwidth]{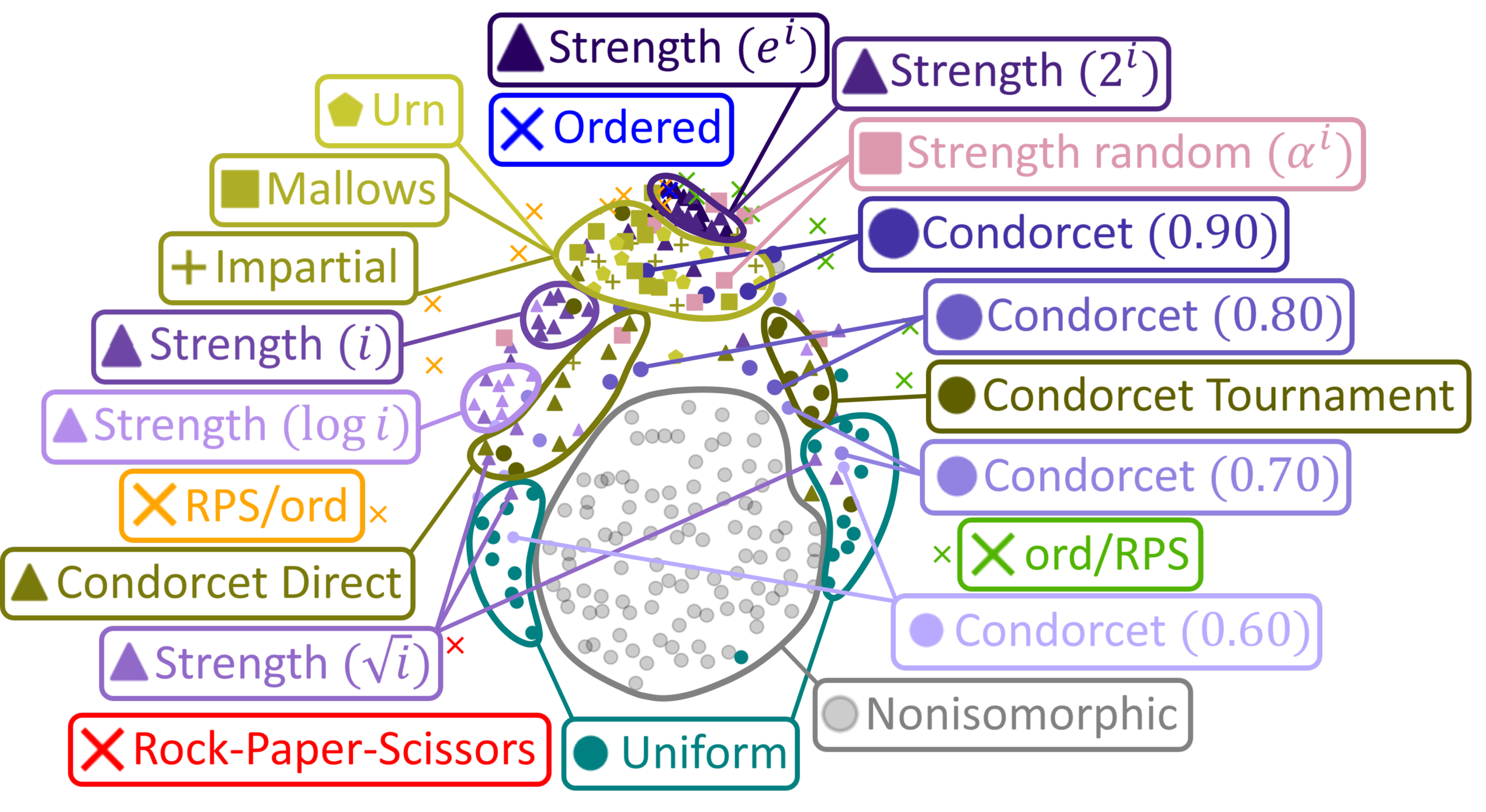}%
    \caption{Graph Edit Distance}
    \label{...}
  \end{subfigure}  
  \begin{subfigure}{0.45\textwidth}
    \includegraphics[width=\textwidth]{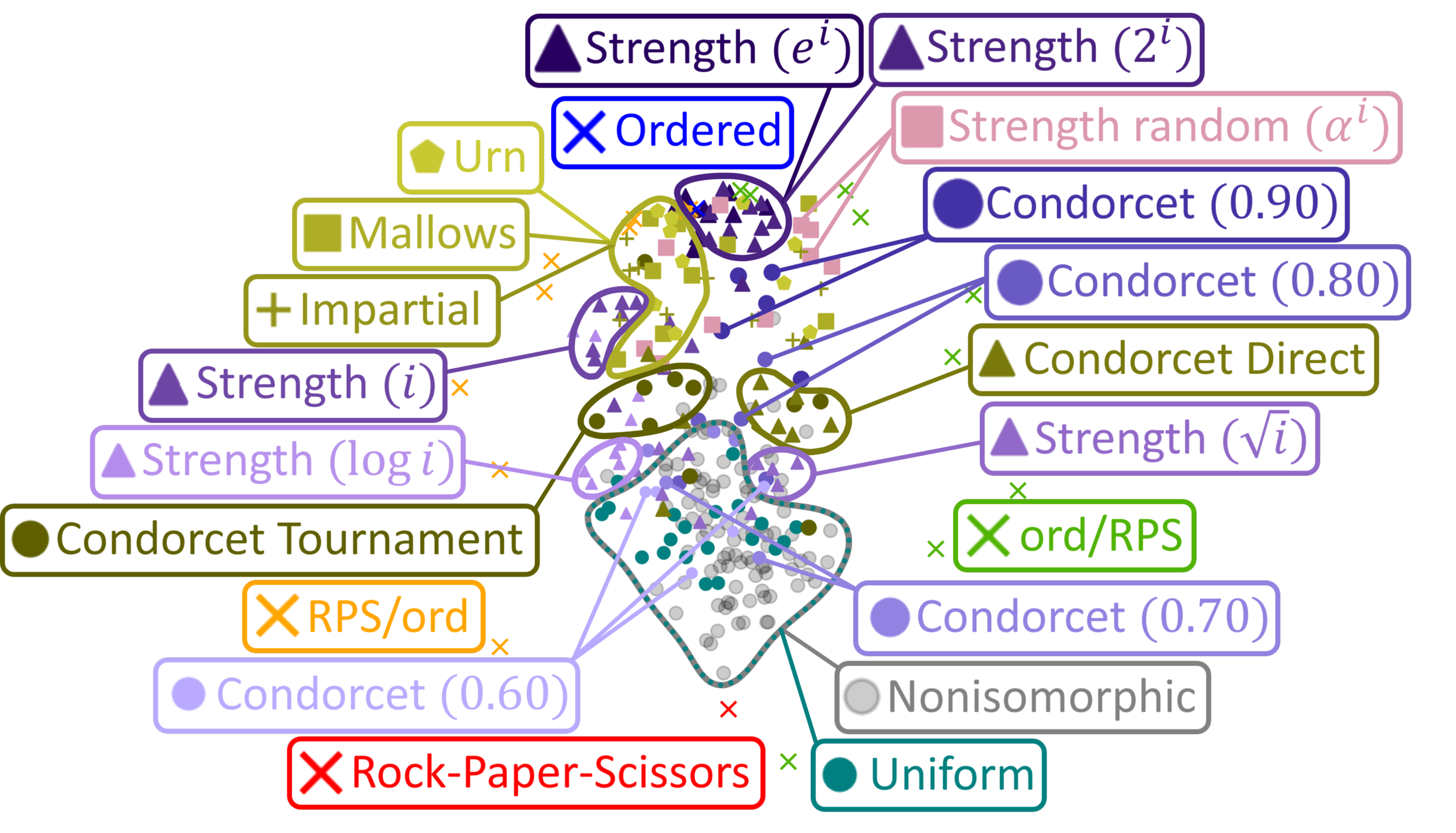}%
    \caption{Katz Distance}
    \label{...}
  \end{subfigure}
  \caption{\label{fig:map:mov} Maps of the MoV dataset.}
\end{figure}

\subsection{Katz Distance}

Our second distance is based on the %
idea %
that two tournaments are similar if Katz centralities of their
players are similar.  For a tournament $T$, we write
$\katz^\downarrow(T)$ to denote the vector of Katz centralities of the
players in $T$, sorted in the nonincreasing order.

\begin{definition}
  Let $T_1$ and $T_2$ be two tournaments over the same number of players.  We
  define their Katz distance as:
   $ d_\katz(T_1,T_2) =
    \ell_1(\katz^\downarrow(T_1),\katz^\downarrow(T_2))$.
\end{definition}

\begin{remark}
  Sorting the vectors of Katz centralities ensures that the Katz
  distance is isomorphic.
\end{remark}

Since Katz centrality can be computed in polynomial time, and we only
compute it $2n$ times to obtain the Katz distance, computing $d_\katz$
can also be done in polynomial time. Katz distance is a pseudodistance
as two different tournaments can have the same sorted Katz vectors
without being isomorphic.

We can define analogous distances using various other centrality
measures and, indeed, we tried nine different ones (see
\Cref{tab:cen-cor} for their list and see \Cref{apdx:centralities} for
their definitions; most of them are also covered in the overview of
\citet{kos-leh-pee-ric-ten-zlo:b:centralities}).  It turned out that
Katz is superior, both because the maps it generates most resemble
those produced using GED and because it is most strongly correlated
with GED, as measured using Pearson and Spearman correlation
coefficients (PCC and SCC, computed for the Size-12 dataset; see \Cref{tab:cen-cor}).
We illustrate the correlation between the Katz distance and GED in
\Cref{fig:size12:ged-katz} (analogous figures for the other
centrality-based distances are in \Cref{apdx:centralities}).

\begin{table}
\bigskip
    \centering
    \begin{tabular}{lcc}
        \toprule
        Centrality measure & PCC & SCC \\
        \midrule
        Katz Centrality & 0.573 & 0.513 \\
        Degree Centrality & 0.558 & 0.493 \\
        Closeness Centrality & 0.440 & 0.388 \\
        Eigenvector Centrality & 0.429 & 0.403 \\
        Laplacian Centrality & 0.422 & 0.378 \\
        Harmonic Centrality & 0.397 & 0.365 \\
        PageRank & 0.375 & 0.367 \\
        Betweenness Centrality & 0.243 & 0.283 \\
        Load Centrality & 0.243 & 0.282 \\
        \bottomrule
    \end{tabular}
    \caption{PCC and SCC between centrality-based
      distances and GED on the Size-12 dataset.}
    \label{tab:cen-cor}
\end{table}

Additionally, in \Cref{fig:uniform:ged-katz} we show how the PCC
between the Katz distance and GED changes as we vary the number of
candidates, while looking at the distance between $T_\Ord$ and
tournaments generated from the uniform model.  While we can only
consider up to 12 players (due to intractability of GED), we see that
this value seems to stabilize. This is a good sign as distance from
$T_\Ord$ is one of the more important features of both the synthetic
and real-life tournaments that we consider (this will become visible
in the maps presented in the following two sections).

\section{Map of Tournaments}

For each of our four datasets and both distances (GED and Katz), we
have prepared its map as follows: For each pair of tournaments we
computed the distance between them (we call these distances
\emph{original}) and, then, for each tournament we found a point on a
2D plane so that the Euclidean distances between the points would
resemble the original ones (we used the MDS
algorithm~\citep{kru:j:mds}). The resulting maps are
in~\Cref{fig:map:all_datasets} (see \Cref{app:25-dataset}
a map with larger tournaments).  We comment on the maps' features
below.

\paragraph{GED Versus Katz.}
While the maps obtained using GED are %
different from
those obtained using the Katz distance,
both seem to provide the same
intuitions. For example, in
Figures~\ref{fig:dataset-12-ged} and \ref{fig:dataset-12-katz}
we see that the bridge tournaments form two
separate clusters in the GED map, but only a single cluster in the
Katz map. However, in both types of maps these tournaments
have similar properties: They are far from $T_\Ord$, located in the
similar part of the tournament space
(given that Katz fails to distinguish the 
tournaments provided by \texttt{nauty}).

\paragraph{Quality of the Embedding.}
For a pair of tournaments $T_1$ and $T_2$ on a given map, we write
$d(T_1,T_2)$ to denote their original distance (either GED or Katz)
and $d_{\euc}(T_1,T_2)$ to denote the Euclidean distance between their
points (in both cases, normalized by the respective distance between
$T_\Ord$ and $T_{\Rps}$). Then, we define the distortion of $T_1$ and
$T_2$ as:
\[
\frac{\max( d(T_1,T_2), d_{\euc}(T_1,T_2) )}{\min( d(T_1,T_2), d_{\euc}(T_1,T_2) )};
\]
the notion was first used to evaluate maps in the map framework
by
\citet{boe-bre-elk-fal-szu:c:frequency-matrices}).

\begin{figure}[t]
  \centering
  \begin{subfigure}{0.24\textwidth}
    \includegraphics[width=\columnwidth]{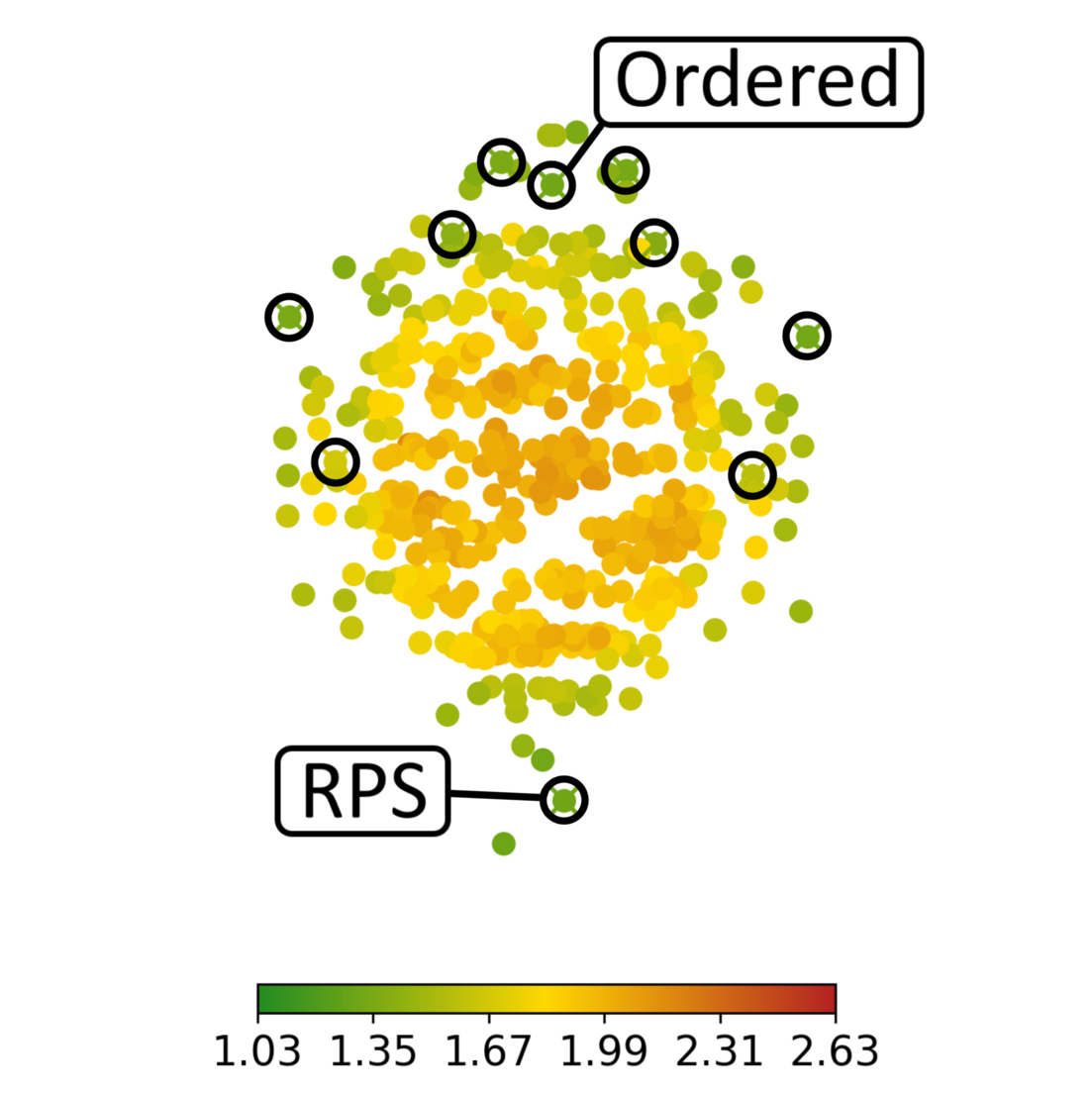}%
    \caption{GED, Size-7 dataset.}
  \end{subfigure}~
  \begin{subfigure}{0.24\textwidth}
    \includegraphics[width=\columnwidth]{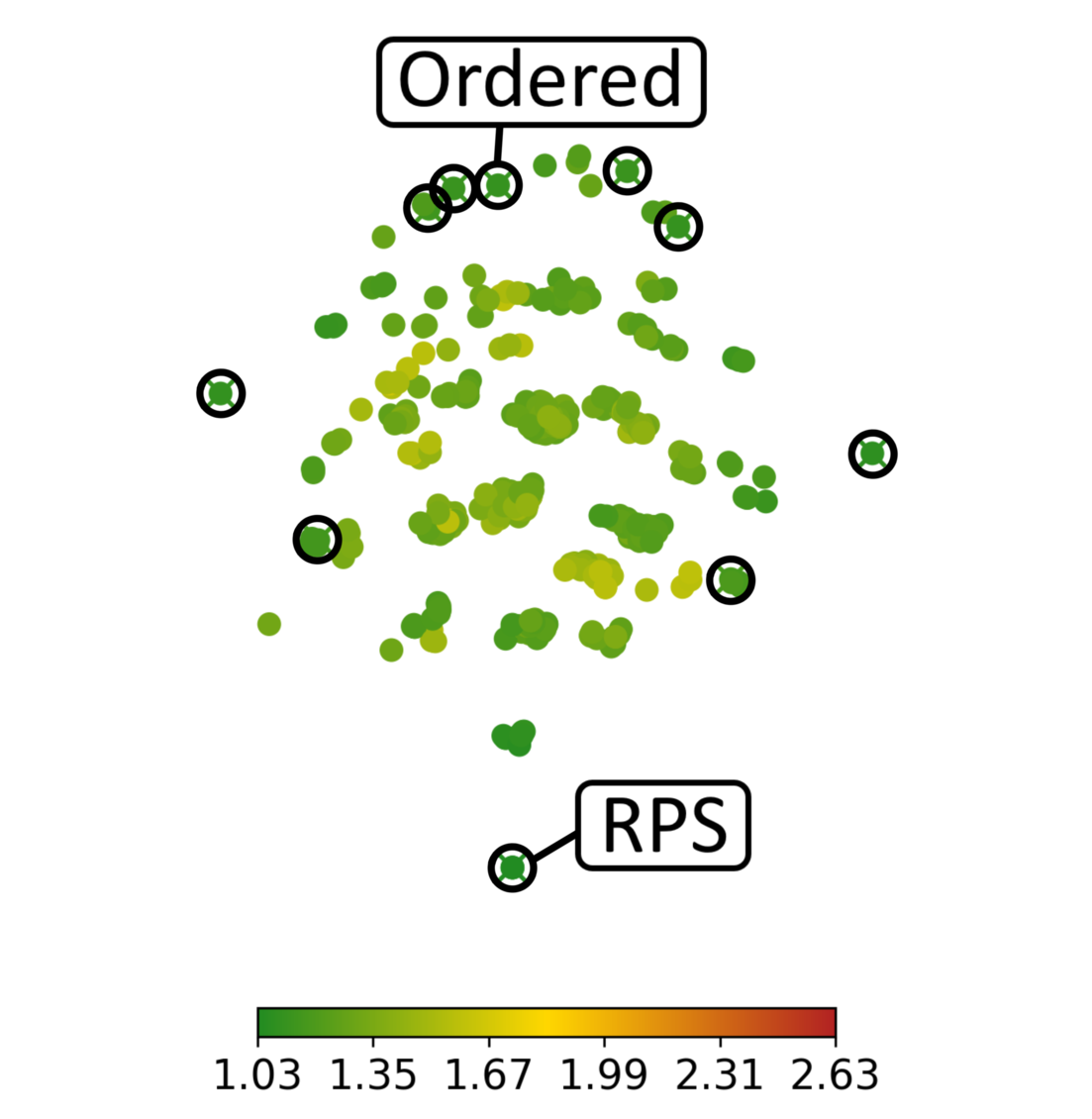}%
    \caption{Katz, Size-7 dataset.}
  \end{subfigure}
  \begin{subfigure}{0.24\textwidth}
    \includegraphics[width=\columnwidth]{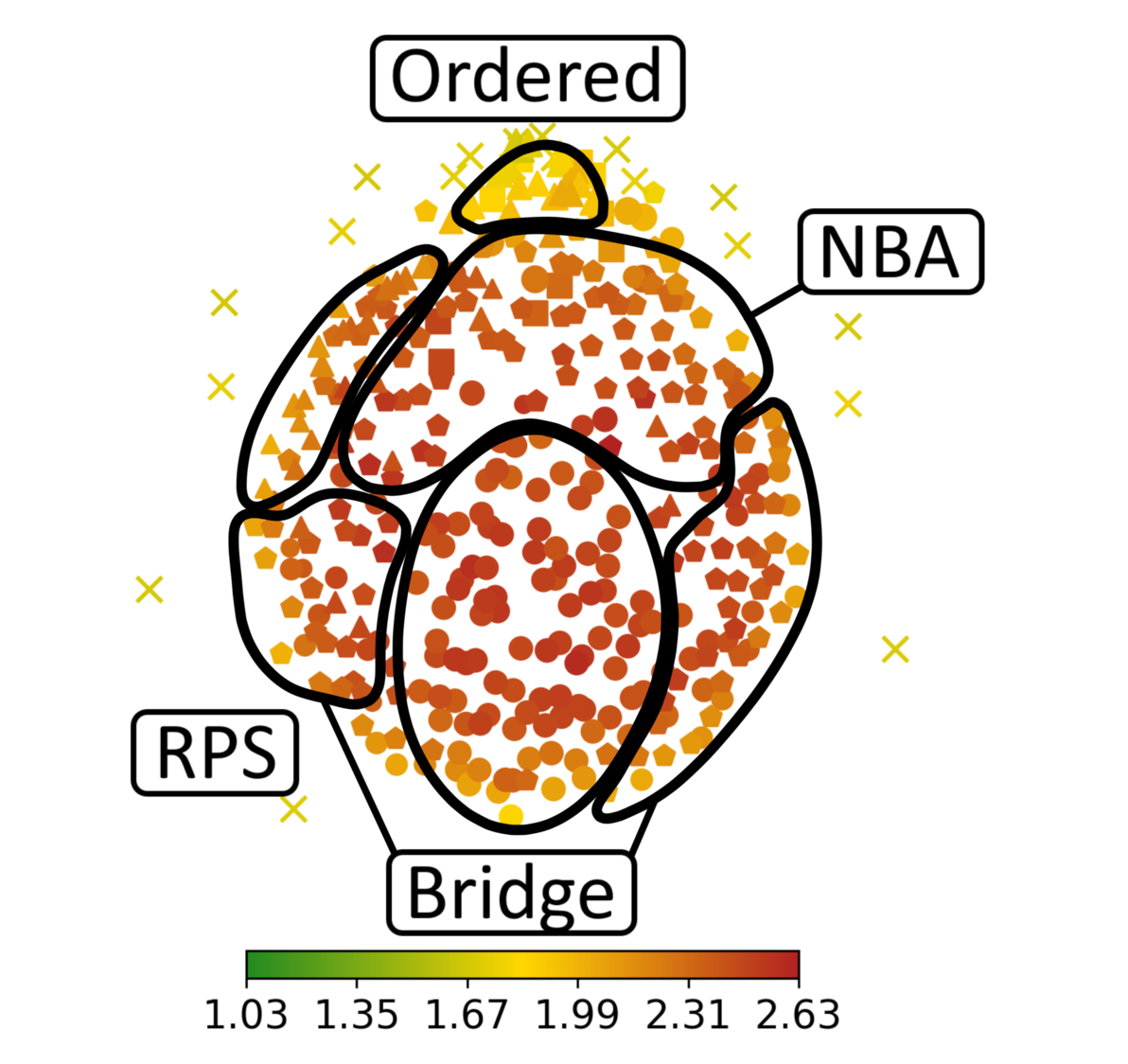}%
    \caption{GED, Size-12 dataset.}
  \end{subfigure}~
  \begin{subfigure}{0.24\textwidth}
    \includegraphics[width=\columnwidth]{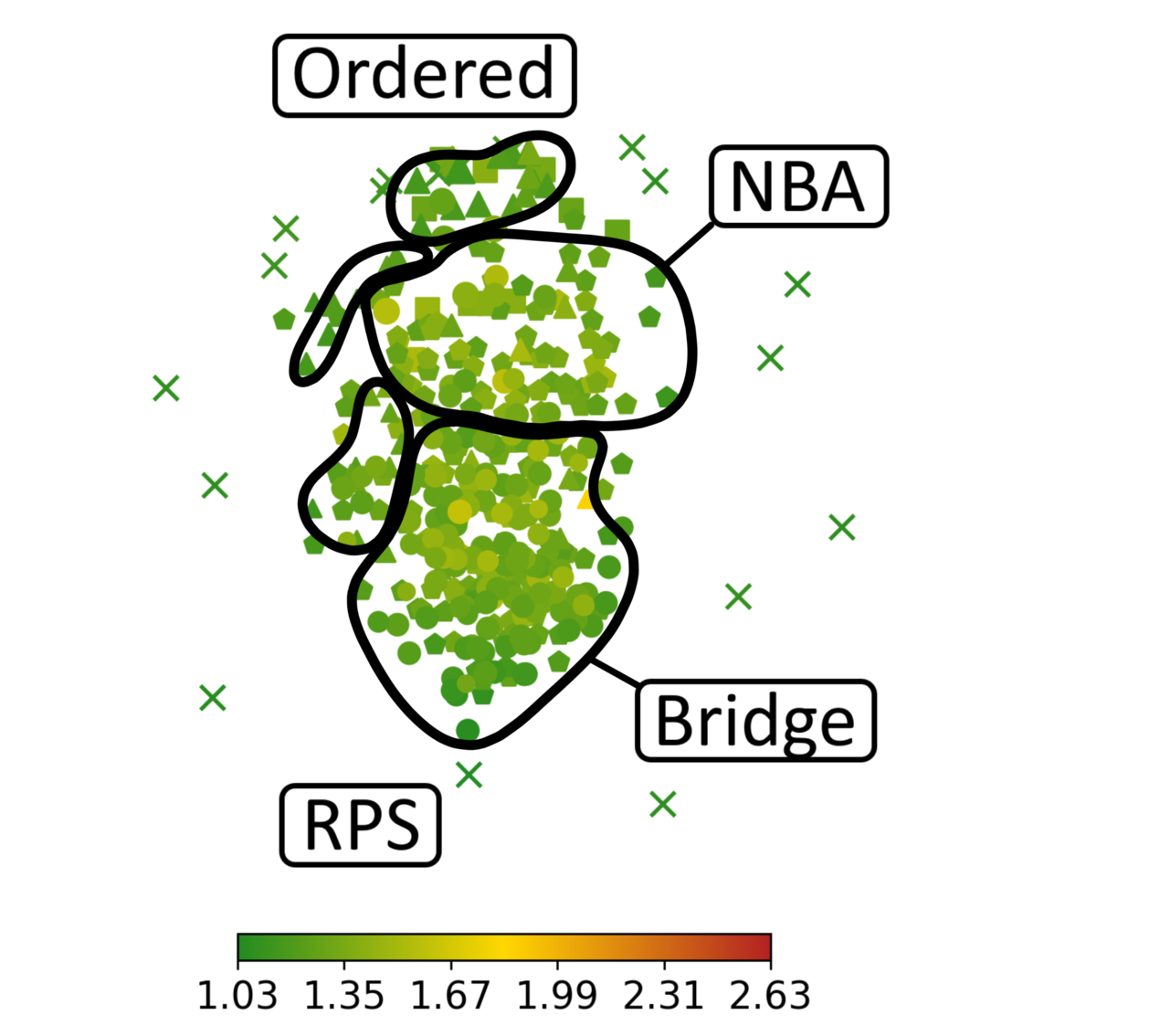}%
    \caption{Katz, Size-12 dataset. }
  \end{subfigure}
  \caption{Distortion of the  the Size-7 and Size-12 maps.}
  \label{fig:distortion-7-12}
\end{figure}

To evaluate the quality of an embedding, for each tournament we
compute its average distortion (i.e., the average over distortions
between this tournament and all the other
ones). In~\Cref{fig:distortion-7-12} %
we present the maps where each point is colored according to its
distortion.
We see that the distortions are similar for both the Size-7 and
Size-12 datasets and, generally, the Katz maps are presented more
accurately. A possible explanation is that GED assumes only a very
limited set of values (e.g., for the Size-12 dataset, all GED
distances are in the set $\{0,1,\ldots, 20\}$), which makes it
difficult to present them in 2D space
(\citet{boe-bre-elk-fal-szu:c:frequency-matrices} achieve much better
distortion values for maps of ordinal elections, but the maps of
stable roommates produced by
\citet{boe-hee-szu:j:map-stable-matching} are similar in this respect
to ours\footnote{Formally,
  \citet{boe-bre-elk-fal-szu:c:frequency-matrices} look at slightly
  different notion than the distortion that we consider, but
  distortion values can be deduced from
  it.}).

\paragraph{Location of Synthetic Tournaments.}
We see that the smaller the $p$ parameter in the Condorcet noise model
or the faster-growing the strength function in the strength models,
the closer are the generated tournaments to $T_\Ord$ in the Size-$12$
maps (Figures~\ref{fig:dataset-12-ged} and \ref{fig:dataset-12-katz}),
which we find quite intuitive.

We note that tournaments generated using the strength models
are closer to the RPS/ord tournaments than to the ord/RPS ones. We
explain this effect on the example of the linear strength function,
$w(i) = i$: We have $12$ players, $v_1, \ldots, v_{12}$, with
strengths $1, \ldots, 12$, respectively. The probability that in the
generated tournament $v_{12}$ beats $v_{11}$ is
$\nicefrac{12}{23} \approx 0.52$. Similarly, the probabilities that
$v_{12}$ beats $v_{10}$, and that $v_{11}$ beats $v_{10}$, are,
respectively, $\nicefrac{12}{22} \approx 0.545$ and
$\nicefrac{11}{21} \approx 0.52$. All these values are quite close to
$0.5$. However, the probability that $v_2$ beats $v_1$ is already
$\nicefrac{2}{3} = 0.66$, which is notably higher. In other words, the
subtournament between the strongest players is more likely to resemble
$T_\Rps$, whereas the subtournament between the weakest ones is more
likely to resemble $T_\Ord$. Since the strongest players are also more
likely to defeat the weakest ones, altogether we are more likely to
get a structure that resembles some tournament from the RPS/ord family
than from the ord/RPS one.

\paragraph{Coverage of the Real-Life Tournaments.}
In Figures~\ref{fig:dataset-12-ged} and~\ref{fig:dataset-12-katz}, we
see that in the areas where NBA and bridge tournaments appear, we also
see synthetically generated tournaments.  Further, aside from the
close vicinity of $T_\Ord$, synthetic tournaments do not appear in
other areas. This suggests that, at least as far as NBA and bridge
goes, our synthetic models are realistic. The MoV dataset (depicted
\Cref{fig:map:mov}) is similar in this respect and, so, the data of
\citet{bri-sch-suk:j:tournaments-mov} covers reasonable parts of the
space of tournaments. On the other hand, most of the tournaments from
the Elections dataset are in the vicinity of $T_\Ord$ (see
\Cref{fig:dataset-election-ged} and
\Cref{fig:dataset-election-katz}). Thus, one should be careful with
using elections to generate tournament data if the purpose is to study
realistic tournaments (and not elections themselves) as one might miss
relevant areas of the tournament space.  On the other hand, our
election-based tournaments are in the same part of the map as the NBA
tournaments and, later on, we will see that this area is particularly
interesting (in the sense that tournaments from this area lead to
diverse results in experiments).

\begin{remark}
  Let us compare the locations of tournaments generated using the
  (election-based) impartial culture model and the (tournament-native)
  uniform model (recall \Cref{rem:ic}).  The former ones appear in the
  top halves of the maps, often quite close to $T_\Ord$, whereas the
  latter ones are placed in the bottom halves, much closer to
  $T_\Rps$. This is surprising as, intuitively, one would expect these
  two models to be very similar. It turns out that the correlations
  between players, introduced in the impartial culture model, are
  strong (recall \Cref{rem:ic}).
\end{remark}

\paragraph{NBA Versus Bridge.} It is striking how the NBA
tournaments are much closer to $T_\Ord$ than the bridge
ones. Intuitively, one would expect that in sports there would be
stronger and weaker teams, and some sort of approximate hierarchy
would be visible in the tournaments. This indeed is so in the NBA
data, but not in bridge. One reason is that unweighted tournaments, as
studied in this paper, are somewhat insufficient to fully capture the
complexity of the bridge league. There, teams not only win or lose,
but also collect points. The final outcome depends on the total number
of collected points. The best teams can play very risky if there is a
chance to collect a large bonus but, in consequence, occasionally they
lose games that they could have won had they played more
conservatively (but, in the end, it is worth it for them). Since such
tournaments happen in real-life, we believe that one should take them
into account in experiments.

\paragraph{Computation Time.} %
Computing the Size-12 map for GED took several days on a server with 2
Intel(R) Xeon(R) Gold 6338 CPU @ 2.00GHz (32 cores with 2 threads in
each CPU), whereas for Katz it took 89 seconds. Times reported in the
following section were obtained on the same machine, but using only a
single thread.

\section{Examples of Experiments Using Maps}
In this section, we use our maps 
to visualize the results of several experiments regarding  winner determination in round-robin
and knockout tournaments. All the experiments use the
Size-12 dataset.

\paragraph{Round-Robin Tournaments.}

We consider the following three rules for selecting winners in
round-robin tournaments (popular in
the literature and %
based on different
principles):
\begin{enumerate}
\item Under the Copeland rule, a score of a given player is equal to
  the number of players that he or she wins against (i.e., his or her
  outdegree). The player with the highest score wins. If there is more
  than one player with the highest score, they all tie as winners.
\item Top cycle of a tournament is the smallest nonempty set of
  players such that each of its members wins against each nonmember.
  The top cycle rule declares as winners all the members of the
  top cycle.
  
\item Under the Slater rule, we %
  find a ranking of players that
  minimizes the number of %
  pairs of  players
  $a, b$ such that $a$ is ahead of $b$ in the ranking,
  but loses to $b$ in the tournament, and declare as %
  the winner %
  the top
  player in this ranking. As there may be several rankings that fulfill
  this condition, the rule can have several tied winners.
\end{enumerate}

\begin{figure}[t]
  \centering
  \begin{subfigure}{0.26\textwidth}
    \includegraphics[width=0.92\columnwidth]{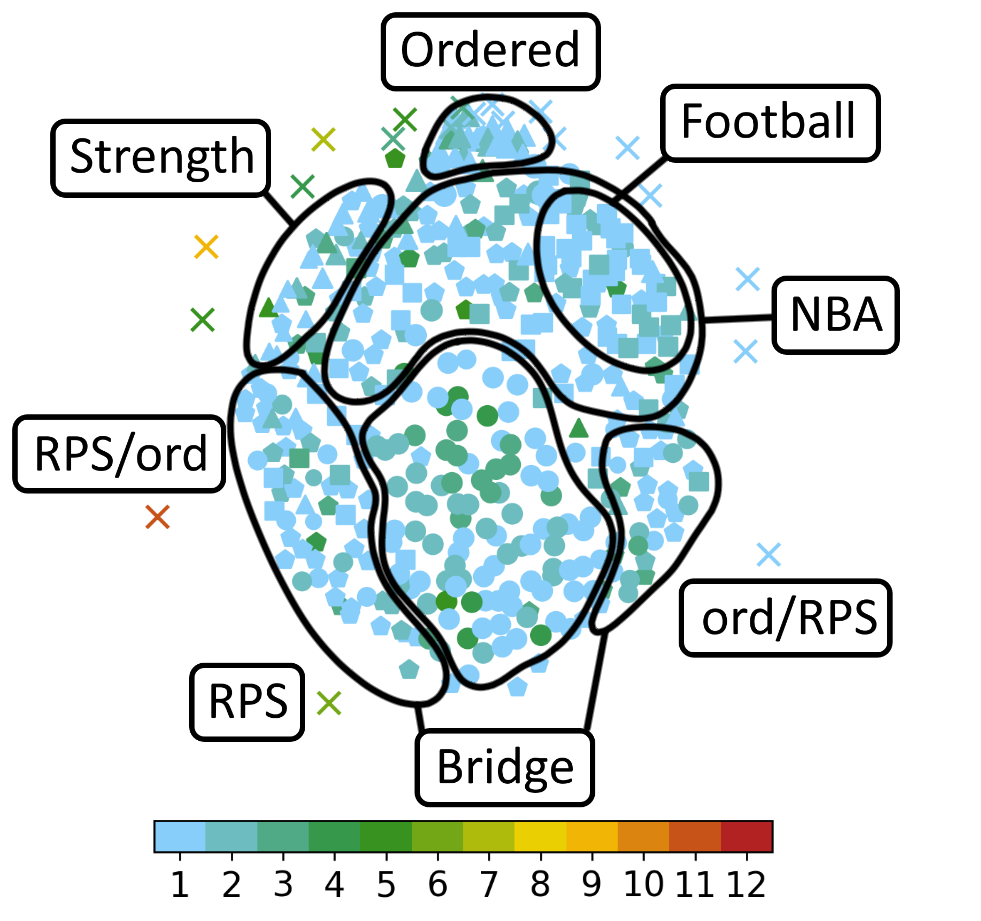}%
    \caption{Copeland winners count.}
    \label{fig:copeland_count}
  \end{subfigure}~
  \begin{subfigure}{0.26\textwidth}
    \includegraphics[width=0.92\columnwidth]{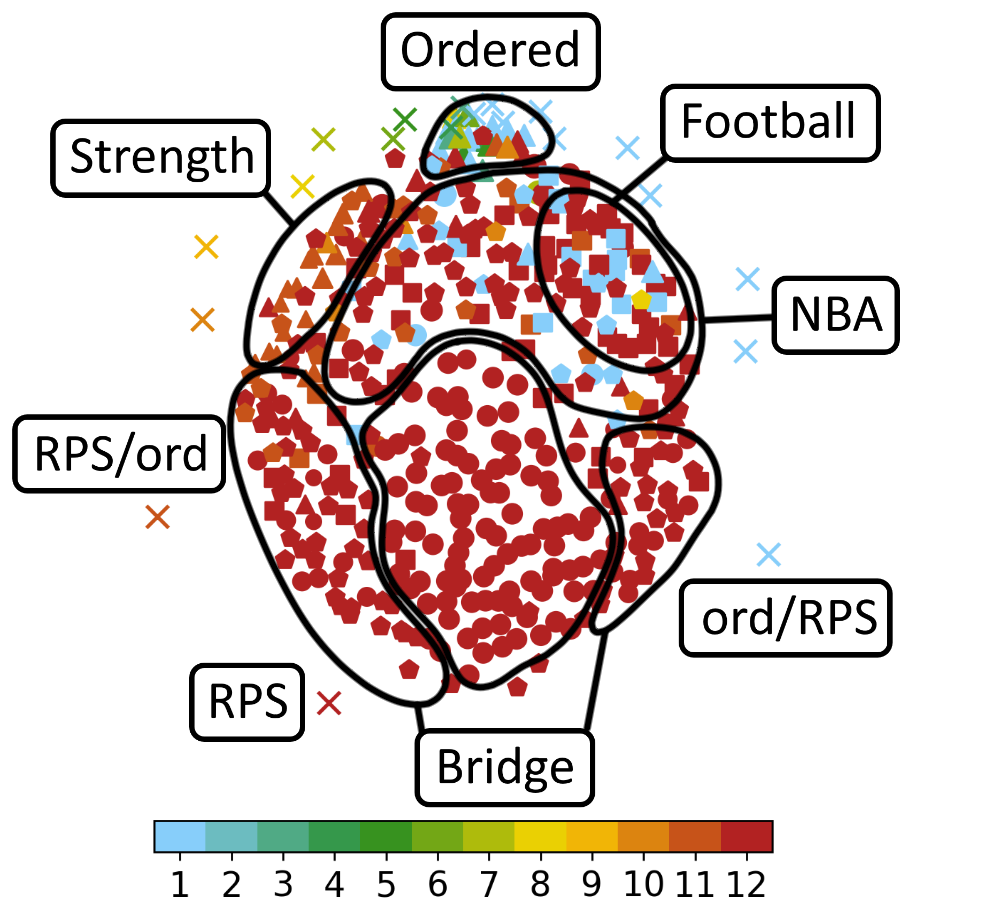}%
    \caption{Top cycle winners count.}
    \label{fig:top_cycle}
  \end{subfigure}~
  \begin{subfigure}{0.24\textwidth}
    \includegraphics[width=\columnwidth, trim={0cm -0.3cm 0cm 0cm}, clip]{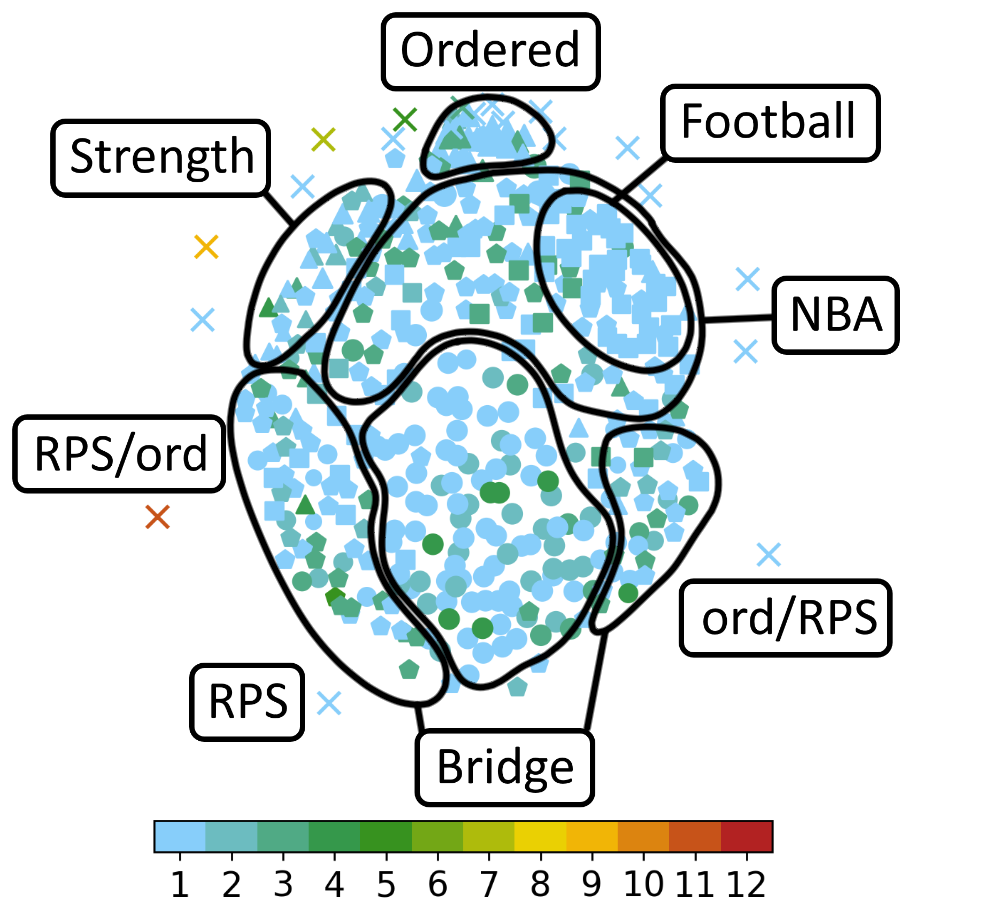}%
    \caption{Slater winners count.}
    \label{fig:slater_count}
  \end{subfigure}~%
  \begin{subfigure}{0.24\textwidth}
    \includegraphics[width=\columnwidth]{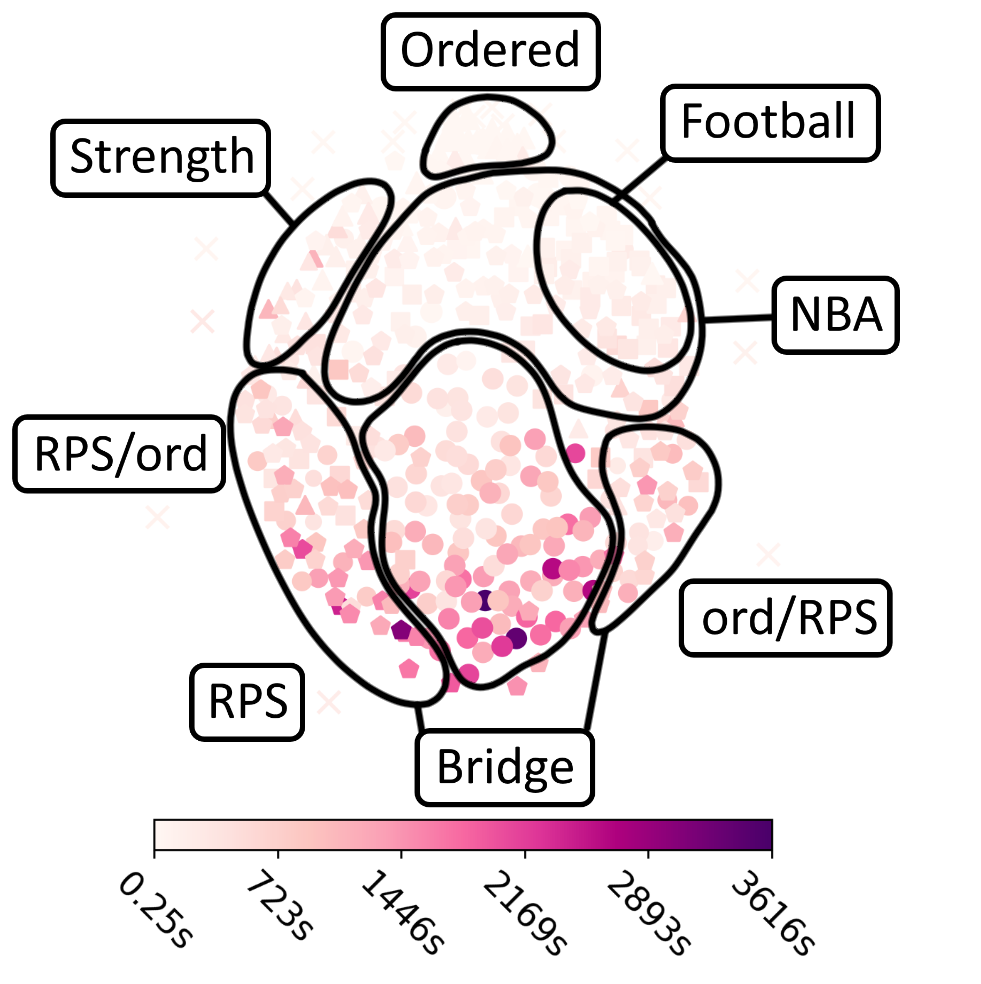}%
    \caption{Slater running time.}
    \label{fig:slater_time}
  \end{subfigure}
  
  \caption{The numbers of winners under (a)~the
    Copeland rule, (b)~top cycle, and (c)~Slater, as well as (d)~the average running time to compute
    Slater winners.}
  \label{fig:voting_rules}
  \bigskip\bigskip
\end{figure}

In Figures~\ref{fig:copeland_count}, \ref{fig:top_cycle}, and
\ref{fig:slater_count} we present maps where each point (tournament)
is colored according to the number of Copeland, top cycle, and Slater
winners, respectively. We see that out of the 368 tournaments in the
dataset, 224 (60\%) have a unique winner under Copeland and 236 (64\%)
have a unique winner under Slater (for each of these
rules there are cases where it has a unique winner but the other one
does not)  %
and no part of the map seems to stand out. %

The top cycle consists of a unique member only for 46 of our
tournaments (12.5\%) and includes all the players in 268 of them
(72\%).  While it was well-known that top cycles often are large, we
found the extent to which this happens in the Size-12 dataset
surprising.  We note that the NBA tournaments, as well as those in
their close vicinity, have top cycles with the most varied
cardinalities.

In~\Cref{fig:slater_time} we show a map where each tournament's color
gives the time needed to compute the Slater rule on this input using
an ILP solver (computing Slater winners is
$\np$-hard~\citep{alo:j:ranking-tournaments,con:c:slater,cha-tho-yeo:j:slater}):
For each tournament we averaged the computation time over 15 attempts
(average standard deviation was 5\% of the running time). We see that
the running time increases together with the distance from $T_\Ord$,
without a clear dependence on how a particular tournament was
generated, except that very structured tournaments (such as $T_\Rps$
and members of the RPS/ord and ord/RPS families) require very little
time (one could also point out that the most demanding instances
belong to the family of ``nonisomorphic'' tournaments generated using
\texttt{nauty}, but these tournaments are also farthest from
$T_\Ord$).

\paragraph{Knockout Tournaments.}
Given a tournament graph $T$, by a knockout tournament we refer to a
rooted, complete binary tree, where each leaf denotes a player from
$T$ and, then, each inner node is labeled with the player from one of
the child nodes that, according to $T$, wins their game against the
player from the other child node.  The player in the root node is the
winner. (The matching of the players to the leaves of the tree is
called the \emph{bracket}). The complexity of choosing a bracket so
that a particular candidate wins was studied in depth in the
literature~\citep{lan-pin-ros-ven-wal:c:seq-majority-winner,vu-alt-sho:c:tournament-fixing,vas:c:fixing-knockout,azi-gas-mac-mat-stu-wal:j:fixing-tournament};
see also the overviews of \citet{vas:b:knockout-tournaments} and
\citet{suk:c:tournaments} for further references (also regarding the
necessary and sufficient conditions for a player to be a possible
winner) and the work on bribery and manipulation in
tournaments~\citep{mat-gol-kla-mun:j:tournaments}.

\begin{figure}[t]
  \centering
  \begin{subfigure}{0.28\textwidth}
    \includegraphics[width=\columnwidth, trim={0.5cm 0cm 0.5cm 0cm}, clip]{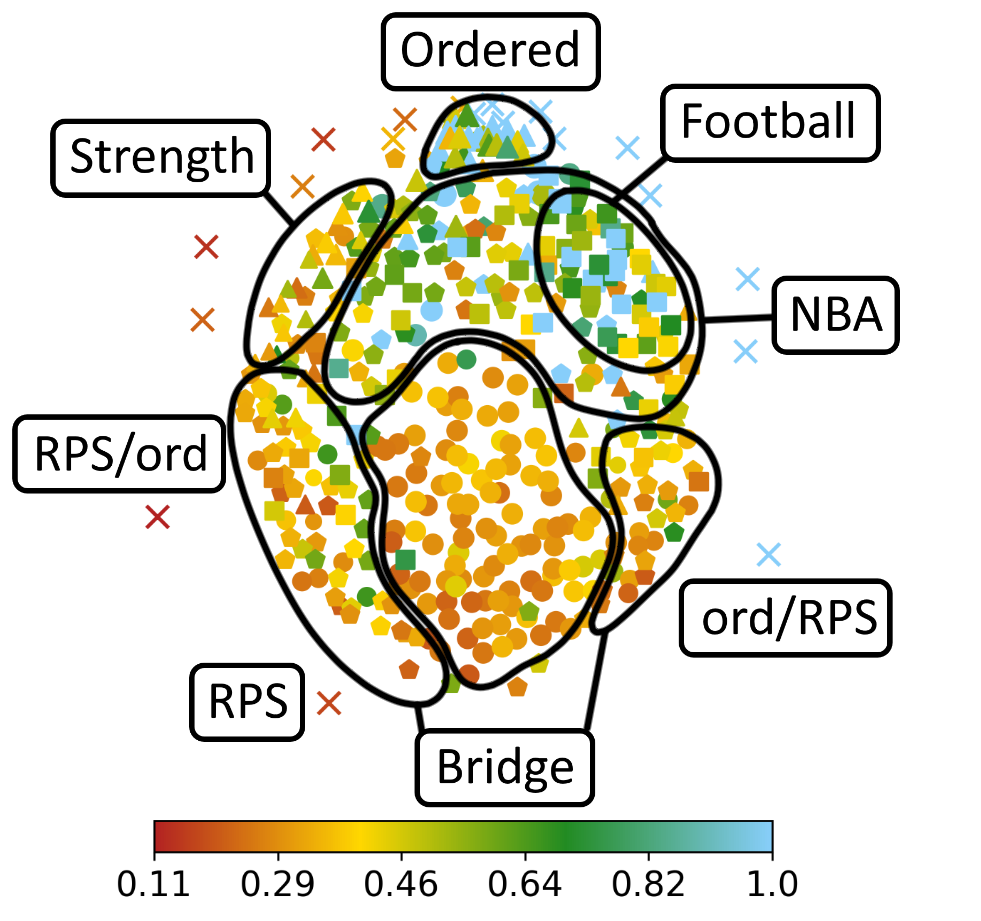}%
    \label{fig:set_probability}
  \end{subfigure}~
  \begin{subfigure}{0.28\textwidth}
    \includegraphics[width=\columnwidth, trim={0.5cm 0cm 0.5cm 0cm}, clip]{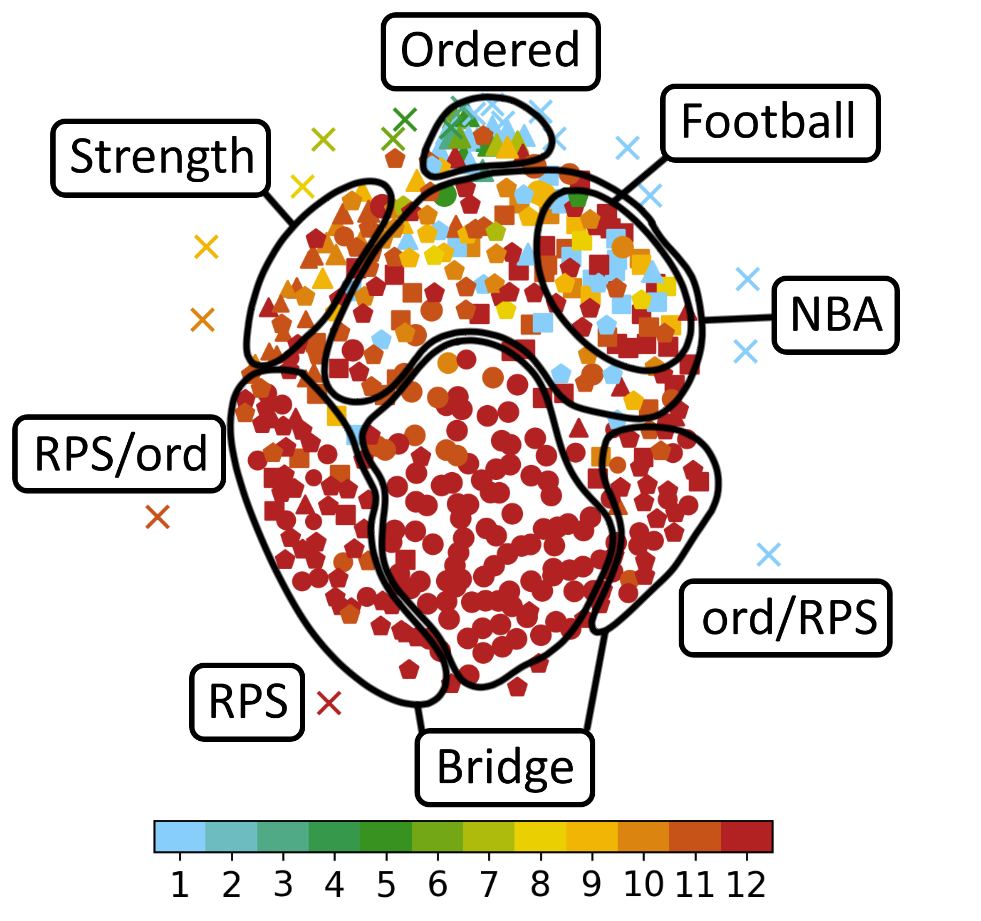}%
    \label{fig:set_possible_winners}
  \end{subfigure}~
  \begin{subfigure}{0.24\textwidth}
    \includegraphics[width=\columnwidth, trim={0.5cm 0cm 0.5cm 0cm},  clip]{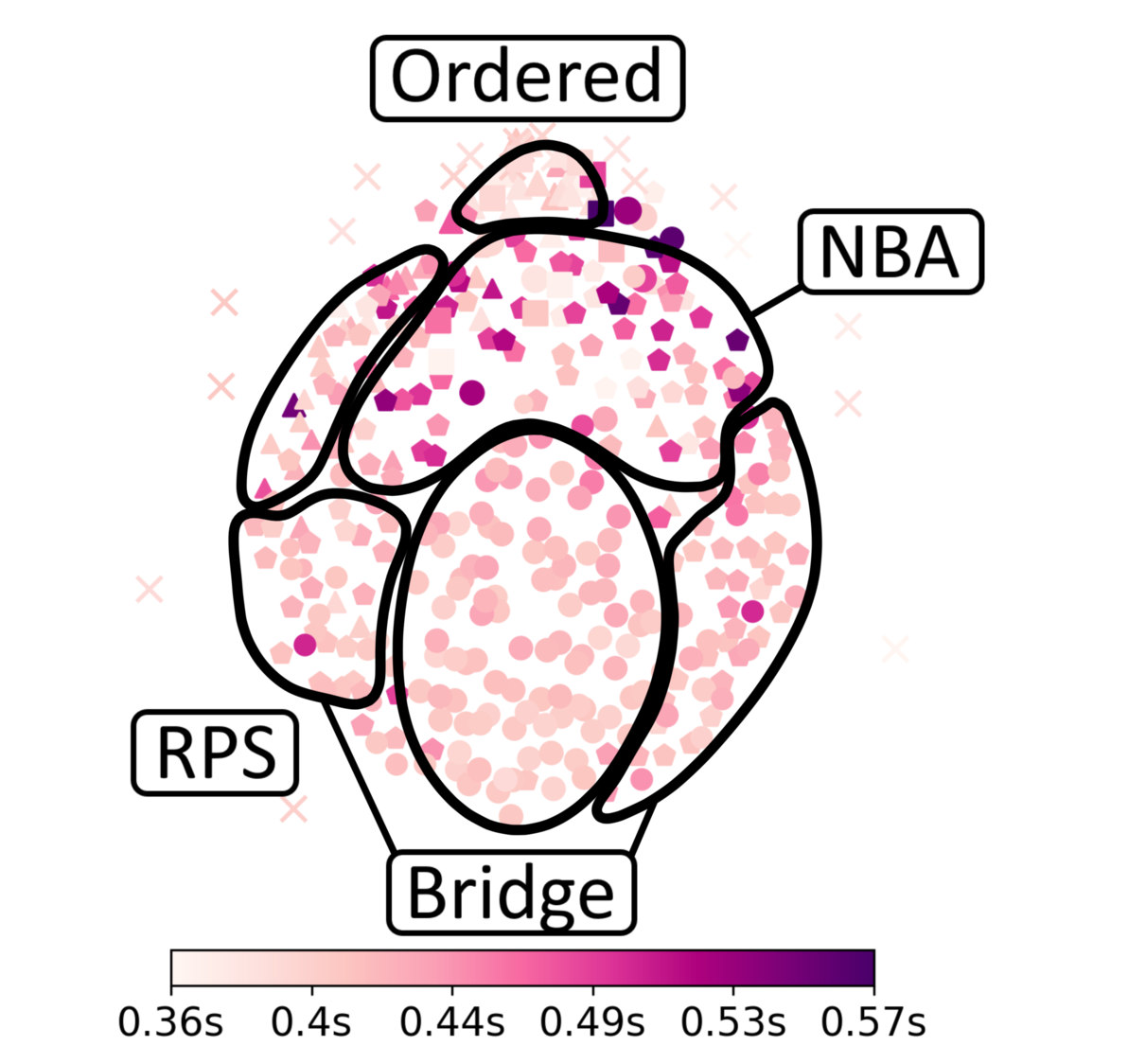}%
    \label{fig:set_time}
  \end{subfigure}
  \bigskip
  
  \caption{The winning probability of a player
    who wins most frequently in a knockout tournament based on the
    given one, under a random bracket (left); The number of possible
    knockout tournament winners (center), and an average time needed
    to check, using an ILP model, which players have brackets under
    which they win (right).}
  \label{fig:knockout}
  \bigskip  \bigskip
\end{figure}

Along the lines of the above-mentioned literature, for each of the
tournaments from the Size-12 dataset we perform the following
experiment: We generate $1000$ random brackets, and we analyze who is
the winner. In~\Cref{fig:knockout} (left), we present the winning
probability of a player who wins most frequently in a given
tournament.  Next, in~\Cref{fig:knockout}~(center) we present the
number of players that may win a knockout tournament based on a given
one (computed using an ILP model), and in~\Cref{fig:knockout}~(right)
the time needed to find these possible winners (averaged over five
attempts).
In the majority of tournaments (i.e., $55\%$ of them),
for each player there is bracket under which he or she wins.
Given these results, it would be interesting to see where real-life
knockout tournaments (such as the tennis ones) appear on the
map. However, to place them there, we would need the results of all
possible matches, and not just those actually played.

While the maps in \Cref{fig:knockout} present different data than
those in the preceding section, their colorings show similar
trends. In particular, the depicted values either seem to be
correlated with the distance from $T_\Ord$ (left) or are most varied in the
NBA area (center and right).

\section{Conclusions and Future Work}

We have adapted the ``map of elections'' framework to the case of
tournament graphs. While doing so,
we have found that direct methods for sampling tournaments seem to
cover the space of those realistic tournaments that we analyzed,
whereas using majority relations of ordinal elections (generated using
typical statistical models) gives tournaments in quite a specific part
of the space.
We obtained two datasets of tournaments, one based on the NBA
basketball league and one based on the Polish bridge leagues.
Results of our experiments were either correlated with the
distance from the ordered tournament $T_\Ord$ or were most varied on
the tournaments in the vicinity of the NBA ones (where also the 
election-based tournaments are located).

As far as future work goes, the most pressing issue is to find a
distance that would be fast to compute, would capture relations
between tournaments well, and could be applied to tournaments of
different sizes. Further, it would be interesting to find statistical
models of tournaments that appear in those parts of the map where our
GED maps only show the tournaments from \texttt{nauty}.

\section{Acknowledgements}
This project has received funding from the European Research Council
  (ERC) under the European Union’s Horizon 2020 research and
  innovation programme (grant agreement No 101002854), from the French
  government under the management of Agence Nationale de la Recherche
  as part of the France 2030 program, reference ANR-23-IACL-0008.

\begin{center}
    \includegraphics[width=3cm]{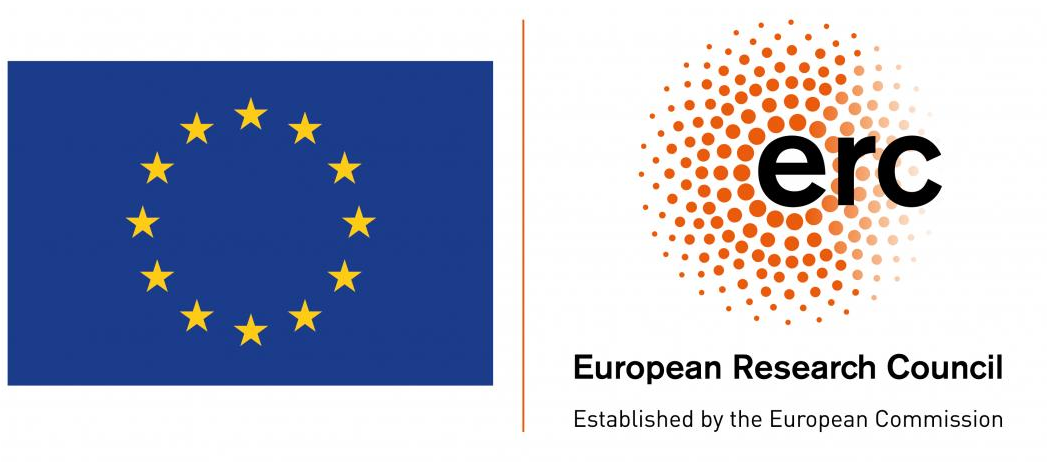}
\end{center}

\bibliography{pragma}

\appendix

\section{Map of the Size-25 Dataset}\label{app:25-dataset}
\begin{figure}[th]
    \centering
    \includegraphics[width=0.45\columnwidth]{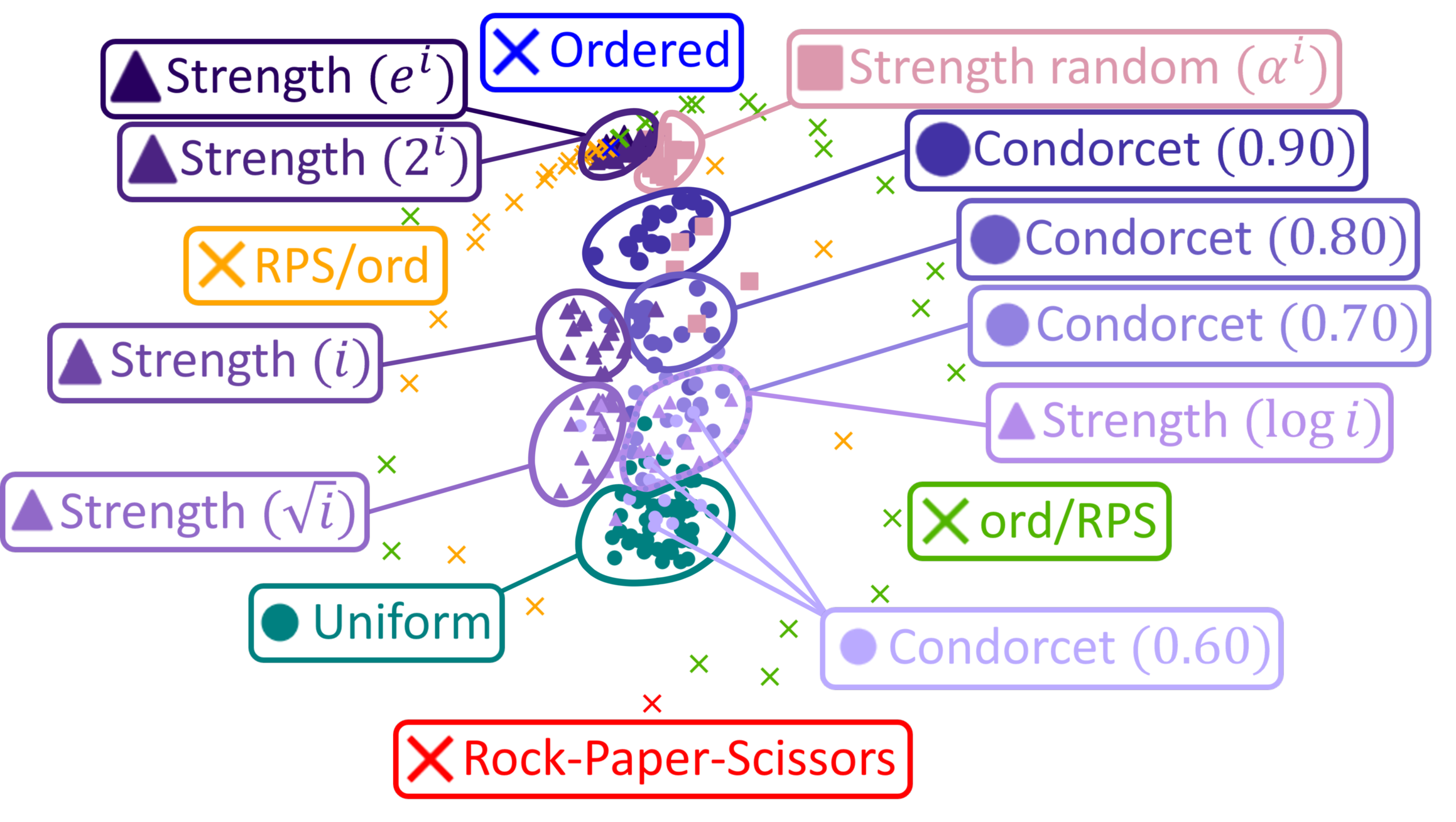}%
    \label{...}
  \caption{\label{fig:map:size-25} Map of the size-25 dataset using Katz distance.}
  \bigskip\bigskip
\end{figure}

In~\Cref{fig:map:size-25} we present the map for size-25 dataset for
Katz distance (similar map for GED would be infeasible). This map
resembles the one for Size-12 dataset, but omits the real-life
tournaments. The main shape of the map is maintained. The biggest
difference is that all instances (with exception for RPS/ord and
ord/RPS) are more clustered, but at the same time they are better
separated.

\section{Centrality-Based Distances}\label{apdx:centralities}

\begin{figure}[t]
  \centering
  \begin{subfigure}{0.33\textwidth}
    \centering
    \includegraphics[width=\columnwidth]{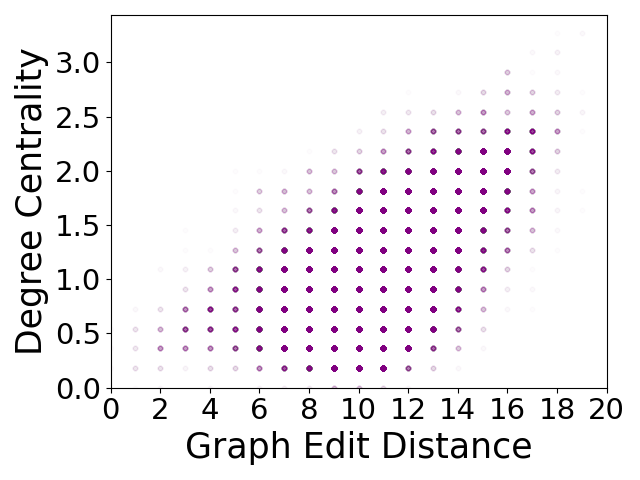}%
    \caption{GED vs Degree Centrality}
    \label{fig:corr:degree}
  \end{subfigure}%
  \hfill
  \begin{subfigure}{0.33\textwidth}
    \centering
    \includegraphics[width=\columnwidth]{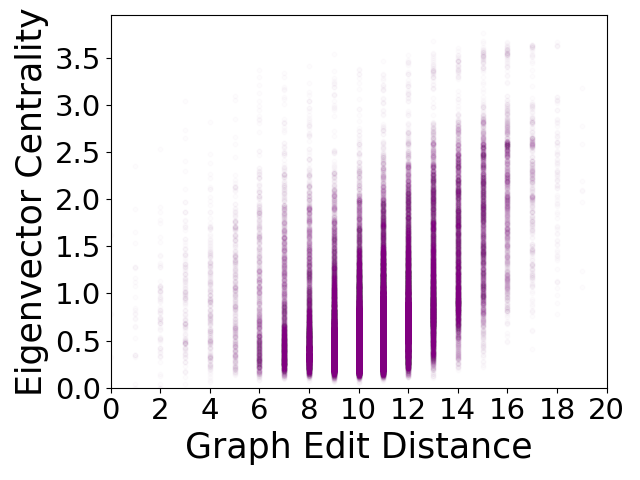}%
    \caption{GED vs Eigenvector Centrality}
    \label{fig:corr:eigen}
  \end{subfigure}%
  \hfill
  \begin{subfigure}{0.33\textwidth}
    \centering
    \includegraphics[width=\columnwidth]{pictures/correlations/corr_ged_blp_katz_cen.png}%
    \caption{GED vs Katz Centrality}
    \label{fig:corr:katz}
  \end{subfigure}%
  \bigskip\bigskip

  \begin{subfigure}{0.33\textwidth}
    \centering
    \includegraphics[width=\columnwidth]{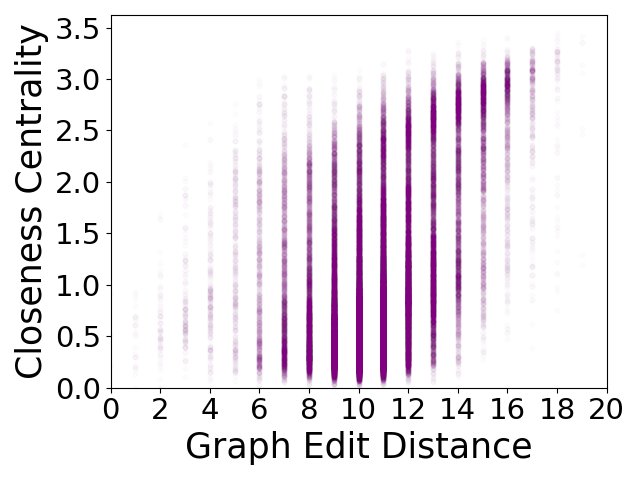}%
    \caption{GED vs Closeness Centrality}
    \label{fig:corr:closeness}
  \end{subfigure}%
  \hfill
  \begin{subfigure}{0.33\textwidth}
    \centering
    \includegraphics[width=\columnwidth]{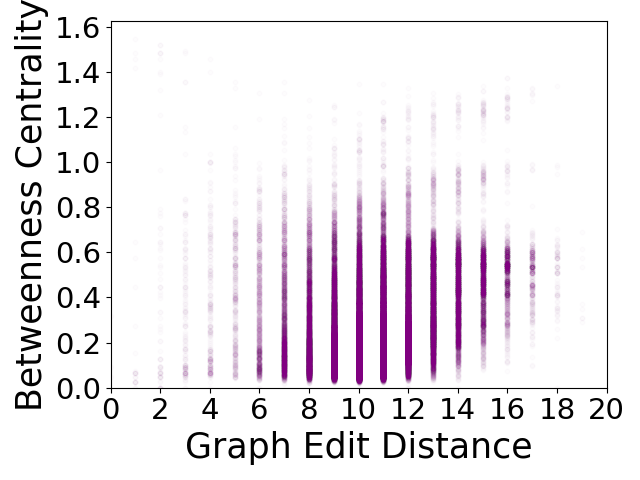}%
    \caption{GED vs Betweenness Centrality}
    \label{fig:corr:betweenness}
  \end{subfigure}%
  \hfill
  \begin{subfigure}{0.33\textwidth}
    \centering
    \includegraphics[width=\columnwidth]{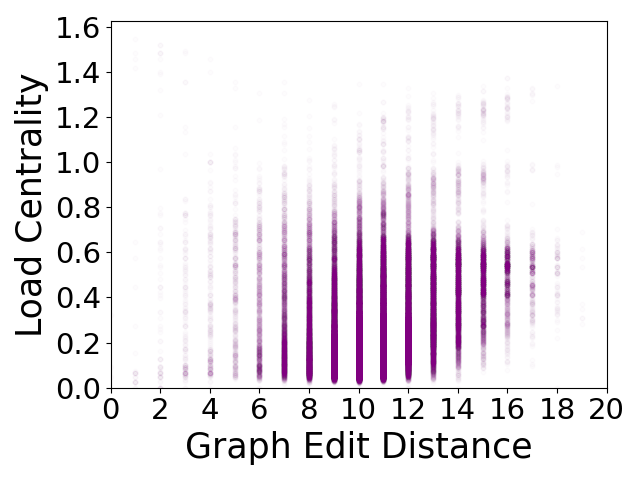}%
    \caption{GED vs Load Centrality}
    \label{fig:corr:load}
  \end{subfigure}%
  \bigskip\bigskip
  
  \begin{subfigure}{0.33\textwidth}
    \centering
    \includegraphics[width=\columnwidth]{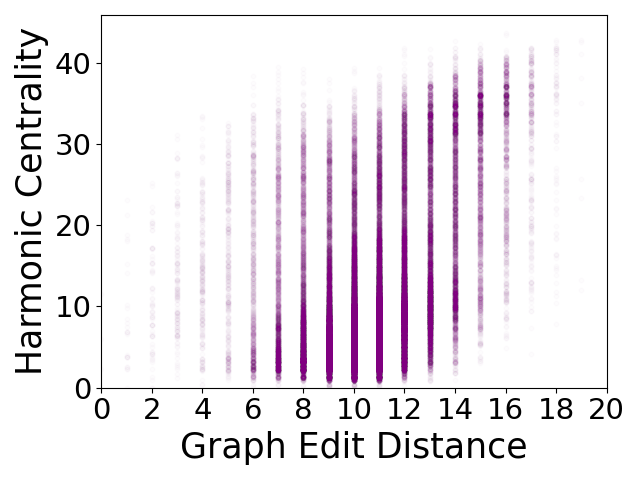}%
    \caption{GED vs Harmonic Centrality}
    \label{fig:corr:harmonic}
  \end{subfigure}%
  \hfill
  \begin{subfigure}{0.33\textwidth}
    \centering
    \includegraphics[width=\columnwidth]{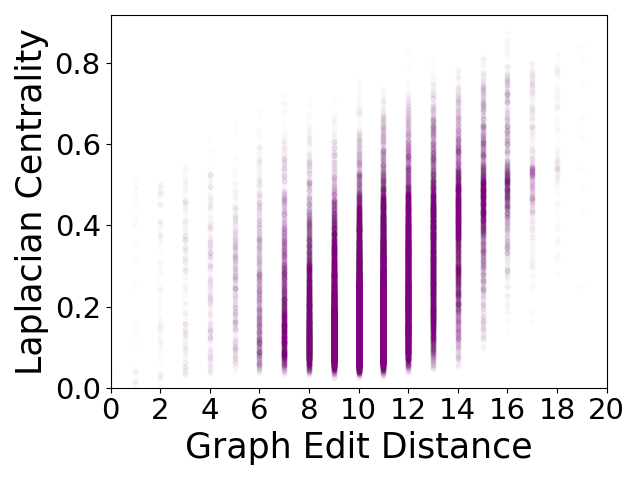}%
    \caption{GED vs Laplacian Centrality}
    \label{fig:corr:laplacian}
  \end{subfigure}%
  \hfill
  \begin{subfigure}{0.33\textwidth}
    \centering
    \includegraphics[width=\columnwidth]{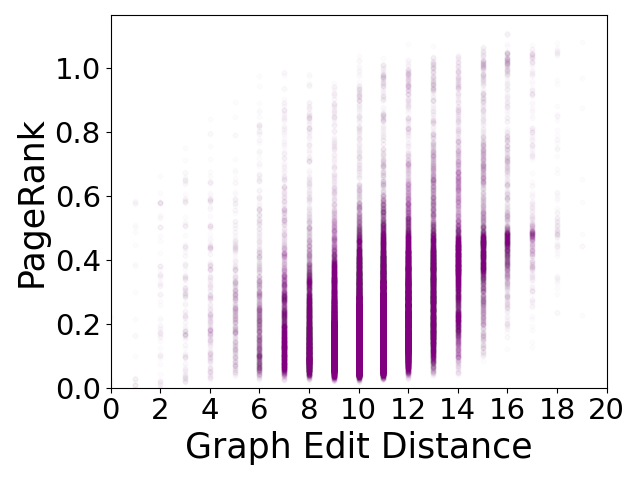}%
    \caption{GED vs Pagerank Centrality}
    \label{fig:corr:pagerank}
  \end{subfigure}
  \bigskip\bigskip

  \caption{\label{fig:cen-pcc}Correlation plots between different centrality-based distances and GED}
\end{figure}

In \Cref{fig:cen-pcc} we visualize the correlation between the nine
centrality measures that we have tested and GED, on the Size-12
dataset.

\subsection{Eigenvector Centrality}
The eigenvector centrality is a remarkably old centrality measure, with the oldest mention we found dating all the way back to 1895 \citeapp{scho-tur:b:eigenvector-centrality}. It is a natural extension of the in-degree centrality, which takes into account not only how many nodes have a direct to the considered node, but also how \textit{important} these nodes are. This means that a node with a lot of incoming connections may have a low score because these connections may be originating from insignificant nodes. Conversely, a node with few in-edges may still be rated very highly if these in-edges come from nodes with a large score. Katz Centrality and Pagerank, which we will explain next, are variants of the eigenvector centrality.

The eigenvector centrality is calculated using the following formula:

\begin{equation}
    \lambda \boldsymbol{c}^T = \boldsymbol{c}^T A
\end{equation}
where $A$ is the adjacency matrix of the graph, $\lambda$ is the largest eigenvalue of the adjacency matrix, and $\boldsymbol{c}$ is the vector of centrality values.

We use the \texttt{networkx} implementation available under the \href{https://networkx.org/documentation/stable/reference/algorithms/generated/networkx.algorithms.centrality.eigenvector_centrality_numpy.html#networkx.algorithms.centrality.eigenvector_centrality_numpy}{\texttt{eigenvector\_centrality\_numpy}} function.

\subsection{Katz Centrality}
Katz centrality stems from the eigenvector centrality and it follows the same philosophy---\textit{A node is important if it is linked to by other important nodes}. One of the problems that eigenvector centrality has is its handling of directed graphs, since each node passes its centrality value only through outgoing edges. This can lead to lots of zeros, despite nodes having many outgoing edges. Katz centrality improves upon this weakness by adding a small bias term so that no node has strictly zero centrality. It was first introduced by \citetapp{kat:j:katz}.

The formula to calculate Katz centrality for a single vertex is as follows:

\begin{equation}
    c_i = \alpha \sum_{j\in V} A_{ij} c_j + \beta
\end{equation}
where $A$ is the adjacency matrix of the graph, $V$ is the set of all vertices, $\alpha$ is a constant constrained by $\alpha < \frac{1}{\lambda_{max}}$, $\beta$ is a constant that controls the initial centrality of a node and $c_i$ are members of the centrality vector.

We use the \texttt{networkx} implementation available under the \href{https://networkx.org/documentation/stable/reference/algorithms/generated/networkx.algorithms.centrality.katz_centrality_numpy.html}{\texttt{katz\_centrality\_numpy}} function.

\subsection{PageRank}
PageRank builds upon Katz centrality even further, by mixing it with the degree centrality. It tries to combat the issue where if a central node in a graph acquires too high of a Katz centrality score, it will redistribute that score to all its neighbours, inflating the scores of a large portion of the nodes. It was invented by \citetapp{pag-bri-mot-win} as an algorithm to rank web pages in the Google search engine. 

Below is the formula used to calculate PageRank for a single vertex:

\begin{equation}
    c_i = \frac{1-d}{n} + d\left(\sum_{j\in V\setminus \{i\}} \frac{c_j}{\sum_k A_{kj}}\right)
\end{equation}
where $A$ is the adjacency matrix of the graph, $V$ is the set of all vertices, $n = |V|$ is the number of all vertices, $d$ is the dampening factor (typically $d=0.85$ \citeapp{bri-pag:j:google}) and $c_i$ are elements of the centrality vector. 

We use the \texttt{networkx} implementation available under the \href{https://networkx.org/documentation/stable/reference/algorithms/generated/networkx.algorithms.link_analysis.pagerank_alg.pagerank.html}{\texttt{pagerank}} function.

\subsection{Betweenness Centrality}
The betweenness centrality tries to measure the degree to which nodes stand between each other. It does so by counting the number of shortest paths passing through a vertex. It was formally defined by \citetapp{fre:j:betweenness-centrality}.

The formula used to calculate betweenness centrality for a single vertex is as follows:

\begin{equation}
    c_v =\sum_{s,t \in V} \frac{\sigma(s, t|v)}{\sigma(s, t)}
\end{equation}
where $\sigma(s, t)$ is the number of shortest paths from $s$ to $t$, $\sigma(s, t|v)$ is the number of shortest paths from $s$ to $t$ passing through $v$ and $c_i$ is an element of the centrality vector.

We use the \texttt{networkx} implementation available under the \href{https://networkx.org/documentation/stable/reference/algorithms/generated/networkx.algorithms.centrality.betweenness_centrality.html}{\texttt{betweenness\_centrality}} function.

\subsection{Load Centrality} 
The load centrality is a measure of the total load or demand on each node in the network. It is very similar to the betweenness centrality, but instead of considering the shortest paths passing through a node, it counts how many \textit{random} walks traverse said node. Load centrality was first proposed by \citetapp{new:j:load-centrality}.

We use the \texttt{networkx} implementation available under the \href{https://networkx.org/documentation/stable/reference/algorithms/generated/networkx.algorithms.centrality.load_centrality.html}{\texttt{load\_centrality}} function.

\subsection{Closeness Centrality}
As the name suggests, closeness centrality measures how close a node is to all other nodes. It does it by calculating the average length of the shortest paths \textit{to} a considered node. Closeness centrality was first defined by \citetapp{bav:j:closeness-centrality}. 

The following formula is used to calculate the closeness centrality values:

\begin{equation}
    c_s = \frac{n - 1}{\sum_{t\in V} d(s,t)},
\end{equation}
where $V$ is the set of vertices in a graph, $n=|V|$ is the number of vertices, $d(s,t)$ is the length of the shortest path from $s$ to $t$ and $c_s$ is the closeness centrality value for the vertex $s$.

We use the \texttt{networkx} implementation available under the \href{https://networkx.org/documentation/stable/reference/algorithms/generated/networkx.algorithms.centrality.closeness_centrality.html}{\texttt{closeness\_centrality}} function.

\subsection{Harmonic Centrality} 
Harmonic centrality measures the harmonic mean of the shortest path distances from a node to all other nodes in the graph. It is closely related to the closeness centrality, with the difference being that instead of a reciprocal of the sum of distances as in the closeness centrality, we take a sum of reciprocals of distances. The idea of using harmonic mean instead of arithmetic was introduced by \citetapp{bea:j:harmonic-mean-centrality} and was later proven by \citetapp{mar-lat:j:harmonic-centrality} to behave better than the arithmetic mean, in particular for infinite graphs.
The harmonic centrality is defined as:

\begin{equation}
    c_t = \sum_{s \in V \setminus \{t\}}\frac{1}{d(s, t)}
\end{equation}
where $V$ is the set of vertices in a graph, $d(s, t)$ is the length of the shortest path from $s$ to $t$ and $c_s$ is the harmonic centrality value for the vertex $s$. 

We use the \texttt{networkx} implementation available under the \href{https://networkx.org/documentation/stable/reference/algorithms/generated/networkx.algorithms.centrality.harmonic_centrality.html}{\texttt{harmonic\_centrality}} function.

\subsection{Laplacian Centrality}
The Laplacian centrality of a node, first proposed by \citetapp{qi-fu-wu-zha:j:laplacian-centrality}, is measured by the drop in the Laplacian energy after deleting the said node from the graph. Laplacian Energy is defined as the sum of the squared eigenvalues of the graph's Laplacian matrix. 

The value of the Laplacian centrality of a vertex is related to the number of 2-walks\footnote{A 2-walk is a walk consisting of two edges} it takes part in. This places it between a fully local centrality measure, like the degree centrality which only takes into account the direct neighbours of a node, and a global measure, like the betweenness centrality, which considers paths throughout the whole graph. 

In equation form we can write:

\begin{equation}
  c_u(u,G) = \frac{(\Delta E)_v}{E_L (G)} = \frac{E_L (G)-E_L (G_v)}{E_L (G)}
\end{equation}
where $C_u(u,G)$ is the centrality value of node $u$, $E_L(G)$ is the Laplacian energy of graph $G$, $E_L(G_v)$ is the Laplacian energy of graph $G$ after deleting node $v$ and:
\begin{align*}
  E_L (G) = \sum_{i=0}^n \lambda_i^2    
\end{align*}
where $\lambda_v$ are the eigenvalues of $G$’s Laplacian matrix.

We use the \texttt{networkx} implementation available under the \href{https://networkx.org/documentation/stable/reference/algorithms/generated/networkx.algorithms.centrality.laplacian_centrality.html}{\texttt{laplacian\_centrality}} function.

\end{document}